\documentclass[a4paper,fleqn,usenatbib]{mnras}

\usepackage[T1]{fontenc}
\usepackage{ae,aecompl}

\usepackage{graphicx}	% Including figure files
\usepackage{amsmath}	% Advanced maths commands
\usepackage{amssymb}	% Extra maths symbols
\title[CMASS DR12 galaxy clustering]{The clustering of galaxies in the SDSS-III Baryon Oscillation Spectroscopic Survey: Modelling the clustering and halo occupation distribution of BOSS CMASS galaxies in the Final Data Release}
\author[S. Rodr\'iguez-Torres et al.]{Sergio A. Rodr\'iguez-Torres$^{1,2,3}$\thanks{email: sergio.rodriguez@uam.es, Campus de Excelencia Internacional UAM/CSIC Scholar}, 
Chia-Hsun Chuang $^{1,4}$\thanks{MultiDark Fellow},  
Francisco Prada$^{1,2,5,6}$,\newauthor  
Hong Guo$^{7,8}$,
Anatoly Klypin$^{9,10}$, 
Peter Behroozi$^{11}$,
Chang Hoon Hahn$^{12}$,\newauthor     
Johan Comparat$^{1,3}$,   
Gustavo Yepes$^3$,
Antonio D. Montero-Dorta$^8$,\newauthor  
Joel R. Brownstein$^8$, 
Claudia  Maraston$^{13}$,
Cameron K. McBride$^{14}$,\newauthor   
Jeremy Tinker$^{12}$,
Stefan Gottl\"ober$^{4}$,    
Ginevra Favole$^{1,2}$,   
Yiping Shu$^8$, \newauthor      
Francisco-Shu Kitaura$^{4}$,
Adam Bolton$^8$, 
Rom\'an Scoccimarro$^{12}$,\newauthor 
Lado Samushia$^{13,15,16}$,
David Schlegel$^{5}$,
Donald P. Schneider$^{17,18}$\newauthor 
\& Daniel Thomas$^{13}$\\
\\
  $^1$ Instituto de F\'isica Te\'orica, (UAM/CSIC), Universidad Aut\'onoma de Madrid, Cantoblanco, E-28049 Madrid, Spain\\
  $^2$ Campus of International Excellence UAM+CSIC, Cantoblanco, E-28049 Madrid, Spain\\
  $^3$ Departamento de F\'isica Te\'orica M8, Universidad Aut\'onoma de Madrid (UAM), Cantoblanco, E-28049, Madrid, Spain\\
  $^4$ Leibniz-Institut f\"ur Astrophysik Potsdam (AIP), Potsdam, Germany\\
  $^5$ Lawrence Berkeley National Laboratory, 1 Cyclotron Road, Berkeley, CA, 94720, USA\\
  $^6$ Instituto de Astrof\'isica de Andaluc\'ia (CSIC), Glorieta de la Astronom\'ia, E-18080 Granada, Spain\\
  $^7$ Shanghai Astronomical Observatory, Chinese Academy of Sciences, Shanghai, 20030, China\\
  $^8$ Department of Physics and Astronomy, University of Utah, 115 South 1400 East, Salt Lake City, UT 84112, USA\\
  $^9$ Astronomy Department, New Mexico State University, Las Cruces, NM, USA\\
  $^{10}$ Severo Ochoa Associate Researcher at the Instituto de F\'isica Te\'orica (UAM/CSIC), Madrid, Spain\\
  $^{11}$ Space Telescope Science Institute, Baltimore, MD 21218, USA \\
  $^{12}$ Center for Cosmology and Particle Physics, Department of Physics, New York University, NY 10003, New York, USA\\
  $^{13}$ Institute of Cosmology \& Gravitation, University of Portsmouth, Dennis Sciama Building, Portsmouth PO1 3FX, UK\\
  $^{14}$ Center for Astrophysics, Harvard University, 60 Garden Street, Cambridge, MA 02138, USA\\ 
  $^{15}$ Department of Physics, Kansas State University, 116 Cardwell Hall, Manhattan, KS 66506, USA\\
  $^{16}$ National Abastumani Astrophysical Observatory, Ilia State University, 2A Kazbegi Ave., GE-1060 Tbilisi, Georgia\\
  $^{17}$ Department of Astronomy and Astrophysics, The Pennsylvania State University, University Park, PA 16802\\
  $^{18}$ Institute for Gravitation and the Cosmos, The Pennsylvania State University, University Park, PA 16802}
\pubyear{2015}
\begin{document}
\label{firstpage}
\pagerange{\pageref{firstpage}--\pageref{lastpage}}
\maketitle
\clearpage
\begin{abstract}
We present a study of the clustering and halo occupation distribution of BOSS CMASS galaxies in the redshift range 0.43 < z < 0.7 drawn from the Final SDSS-III  Data Release. We compare the BOSS results with the predictions of a Halo Abundance Matching (HAM) clustering model that assigns galaxies to dark matter halos selected from the large \textsc{BigMultiDark} $N$-body simulation of a flat $\Lambda$CDM Planck cosmology. We compare the observational data with the simulated ones on a light-cone constructed from  20 subsequent outputs of the simulation. Observational effects such as incompleteness, geometry, veto masks and fiber collisions are included in the model, which reproduces within 1-$\sigma$ errors the observed monopole of the 2-point correlation function at all relevant scales: from the smallest scales, 0.5 $h^{-1}$ Mpc, up to scales beyond the Baryonic Acoustic Oscillation feature. This model also agrees remarkably well with the BOSS galaxy power spectrum (up to $k\sim1$ $h$ Mpc$^{-1}$), and the Three-point correlation function. The quadrupole of the correlation function presents some tensions with observations. We discuss possible causes that can explain this disagreement, including target selection effects. Overall, the standard HAM model describes remarkably well the clustering statistics of the CMASS sample. We compare the stellar to halo mass relation for the CMASS sample measured using weak lensing in the CFHT Stripe 82 Survey with the prediction of our clustering model, and find a good agreement within 1-$\sigma$. The \textsc{BigMD-BOSS} light-cone including properties of BOSS galaxies and halo properties is made publicly available.
\end{abstract}
\begin{keywords}
Cosmology:large-scale structure of Universe - galaxies: abundances - galaxies: halos - methods: numerical
\end{keywords}
%%%%%%%%%%%%%%%%%%%%%%%%%%%%%%%%%%%%%%%%%%%%%%%%%%
%%%%%%%%%%%%%%%%% BODY OF PAPER %%%%%%%%%%%%%%%%%%

\section{Introduction}

One of the major goals in cosmology is to explain the formation of the large-scale structure of the Universe. However, the main ingredient that drives  this evolution -- the dark matter -- can only be probed using the distribution of galaxies, and galaxies are biased tracers of the matter field. This makes this study challenging. In the last twenty years, vast amounts of observational data have been obtained, improving each time the precision of the large-scale structure measurements and demanding ever more accurate theoretical models. In fact, one of the strongest arguments that we understand how the large-scale structure forms and evolves is our ability to reproduce the galaxy clustering through cosmic time, starting from the primordial Gaussian perturbations. During the last decade, surveys such as the Sloan Digital Sky Survey \citep[SDSS-I/II/III;][]{york2000,eisenstein2011}, have made it possible to determine the clustering of galaxy populations at scales out to tens of Mpc and beyond with reasonable accuracy. 

The Baryon Oscillations Spectroscopic Survey  \citep[BOSS;][]{dawson2013} Data Release 12  \citep[DR12;][]{alam2015} provides redshift of 1.5 million massive galaxies in 10,000 deg$^2$ area of the sky and for redshifts in the range $0.15$ and $0.75$. BOSS DR12 has an effective volume seven times larger than that of the SDSS-I/II project. These data provide us with a sufficiently statistical sample to examine our theoretical predictions over a range of scales. 

In order to compare the $\Lambda$CDM model and the observational data, it is necessary to link the galaxy and the dark matter distributions. There are a number of methods to assign galaxies to the dark matter. State-of-the-art hydrodynamical simulations, that include detailed galaxy formation descriptions, are computationally unaffordable for the volumes considered here  \citep[e.g.,][]{Vogelsberger2014,schaye2015}, and indeed, there are no large samples of simulated galaxies that can be used to match BOSS. Semi-analytic models (SAMs) are less computationally consuming methods to populate dark matter halos with galaxies  \citep[e.g.,][]{knebe2015}. These models incorporate some physics of galaxy formation. 

The most popular models are based on the statistical relations between galaxies and dark matter halos. One of the most used models is the halo occupation distribution \citep[HOD; e.g.,][]{jing1998,peacock2000,berlind2002,zheng2005,leauthaud2012,guo2014}. The main component of the HOD is the probability, $P(N|M_h)$, that a halo of virial mass $M_h$ hosts $N$ galaxies with some specified properties. These models have several parameters which allow one to match the observed clustering. 

The model known as the Halo Abundance Matching \citep[HAM;][]{kravtsov2004,conroy2006,behroozi2010,guo2010,trujillo2011,nuza2013,reddick2013} connects observed galaxies to simulated dark matter halos and subhalos by requiring a correspondence between the luminosity or stellar mass and a halo property. The assumption of this model is that more luminous (massive) galaxies are hosted by more massive halos. However, this relation is not a one to one relation because there is a physically motivated scatter between galaxies and dark matter halos \citep[e.g.,][]{shu2012}. By construction, the method reproduces the observed luminosity function, LF (or stellar mass function, SMF). HAM relates the luminosity function (stellar mass function) of an observed sample with the distribution of halos in a $N$-body simulation. The implemented assignment requires that one works with complete samples in luminosity (stellar mass) or have a precise knowledge of the incompleteness as a function of the luminosity (stellar mass) of the galaxy sample. Luminous red galaxies (LRG) are the most massive galaxies in the universe and they represent the high-mass end of the stellar mass function. This feature makes this population of galaxies an excellent group to be reproduced with the abundance matching. 

In this paper, we compare the clustering of the BOSS CMASS DR12 sample with predictions from $N$-body simulations. We use an abundance matching to populate the dark matter halos of the \textsc{BigMultiDark} Planck simulation \citep[\textsc{BigMDPL};][]{klypin2014}. In order to include systematic effects from the survey, as well as the proper evolution of the clustering, we construct light-cones which reproduce the angular selection function, the radial selection function and the clustering of the monopole in configuration space. To generate these catalogues we developed the SUrvey GenerAtoR code (\textsc{SUGAR}). Once the HAM and the light-cone are applied, we compute the predictions of our model for 2-point statistics and the Three-point correlation function. We also present the prediction of the stellar to halo mass relation and its intrinsic scatter compared to lensing measurements. The HAM, the \textsc{BigMDPL} and the methodology to produce light-cone played a key role in the construction of the \textsc{MultiDark patchy} BOSS DR12 mocks \citep[\textsc{md-patchy} mocks][companion paper]{kitaura2015}.

In order to have a good estimation of the uncertainties in this work, we use 100 \textsc{md-patchy} mocks. These mocks are produced using five boxes at different redshifts that are created with the \textsc{patchy}-code \cite[]{kitaura2014}. The \textsc{patchy}-code can be decomposed into two parts: 1) computing approximate dark matter density field, and 2) populating galaxies from dark matter density field with the biasing model. The dark matter density field is estimated using Augmented Lagrangian Perturbation Theory \citep[ALPT;][]{Kitaura2013} which combines the second order perturbation theory \citep[2LPT; see e.g.,][]{buchert1994,bouchet1995,catelan1995} and spherical collapse approximation \citep[see][]{bernardeau1994,mohayaee2006,neyrinck2013}. The biasing model includes deterministic bias and stochastic bias \citep[for details see][]{kitaura2014}. The velocity field is constructed based on the displacement field of dark matter particles. The modelling of finger-of-god has also been taken into account statistically. The \textsc{md-patchy} mocks are constructed based on the \textsc{BigMD} simulation with the same cosmology used in this work. The mocks match the clustering of the galaxy catalogues for each redshift bin \citep[see][companion paper, for details]{kitaura2015}. The \textsc{BigMultiDark} light-cone catalogues of BOSS CMASS galaxies in the Final DR12 (hereafter \textsc{BigMD-BOSS light-cone}) presented in this work are publicly available.

This paper is structured as follows: sections \ref{sec:boss} and \ref{sec:bigmd} describe the SDSS-III/BOSS CMASS galaxy sample and the \textsc{BigMDPL} $N$-body cosmological simulations used in this work.  In section \ref{sec:sugar}, we provide details on different observational effects and briefly describe the \textsc{SUGAR} code. Section \ref{sec:ham} presents the main ingredients of the HAM modelling of the CMASS galaxy clustering. A comparison of our results to observation is shown in section \ref{sec:result}. Subsequently, we discuss the principal results in section \ref{sec:discussion}. Finally, in section \ref{sec:summary}, we present a summary of our work. For all results in this work, we use the cosmological parameters  $\Omega_m=0.307$, $\Omega_B=0.048$, $\Omega_\Lambda=0.693$.

\section{SDSS-III/BOSS CMASS SAMPLE}
\label{sec:boss}

The Baryon Oscillations Spectroscopic Survey\footnote{http://http://skyserver.sdss.org/dr12/en/home.aspx}  \citep[BOSS;][]{dawson2013,bolton2012} is part of the SDSS-III program \citep{eisenstein2011}. The project used the 2.5 m aperture Sloan Foundation Telescope at Apache Point Observatory \citep{gunn2006}. The telescope used a drift-scanning mosaic CCD camera \citep{gunn1998} with five colour-bands, $u,g,r,i,z$ \citep{fukugita1996}. Spectra are obtained using the double-armed BOSS spectrographs, which are significantly upgraded from those used by SDSS I/II, covering the wavelength range $3600-10000 \textup{\AA}$ with a resolving power of 1500 to 2600 \citep{smee2013}. BOSS provides redshift for 1.5 million galaxies in 10,000 deg$^2$ divided into two samples: LOWZ and CMASS. The LOWZ galaxies are selected to be the brightest and reddest of the low-redshift galaxy population ($z\lesssim 0.4$), extending the SDSS I/II LRGs. The CMASS target selection is designed to isolate galaxies at higher redshift ($z\gtrsim 0.4$), most of them being also luminous red galaxies.

In the present paper, we focus on the CMASS DR12 North Galactic Cap (NGC) sample. Galaxies are selected from SDSS DR8 imaging \citep{aihara2011} according to a series of colour cuts designed to obtain a sample with approximately ``constant stellar mass'' \citep{reid2015}. The following photometric cuts are applied:
\begin{align}
  \label{eq:cut1}
  17.5<i_{cmod}<19.9& \\
  r_{mod}-i_{mod}<2& \\
 d_\perp>0.55& \\
 i_{fib2}<21.5& \\
 i_{cmod}<19.86+1.6(d_\perp-0.8)&
\end{align}
where $i$ and $r$ indicate magnitudes, $i_{fib2}$ is the $i$-band magnitude within a $2''$ aperture. All magnitudes are corrected for Galactic extinction \citep[via  the][dust maps]{schlegel1998}. The subscript $_{mod}$ denotes the ``model'' magnitudes and the subscript $_{cmod}$ refer to the ``cmodel'' magnitudes. The model magnitudes represent the best fit of the DeVaucouleurs and exponential profile in the $r$-band \citep{stoughton2002} and the cmodel magnitudes denote the best-fitting linear combination of the exponential and DeVaucouleurs models \citep{abazajian2004}. $d_\perp$ is defined as
\begin{equation}
  \label{eq:dperp}
  d_\perp=r_{mod}-i_{mod}-(g_{mod}-r_{mod})/8.0.
\end{equation}

Star-galaxy separation is performed on the CMASS targets via:
\begin{align}
  \label{eq:cut2}
  i_{psf}-i_{mod}>0.2+0.2(20.0-i_{mod})&\\
z_{psf}-z_{mod}>9.125-0.46z_{mod},&
\end{align}
The subscript ``psf'' refers to Point Spread Function magnitudes. CMASS sample contains galaxies with redshift $z>0.4$, having the peak of the number density at $z\approx0.5$. We will concentrate our analysis in the redshift range $0.43<z<0.7$ for this sample.

BOSS sample is corrected for redshift failures and fiber collisions. In the following sections, we will use the same weights given in \citet{anderson2014} in order to correct the clustering signal affected by these systematics \citep{ross2012}. The total weight for a galaxy is given by:
\begin{equation}
  \label{eq:weight}
  w_g = w_{star} w_{see} (w_{zf} + w_{cp} - 1).
\end{equation}
In equation \eqref{eq:weight}, $w_{zf}$ denotes the redshift failure weight and $w_{cp}$ represents the close pair weight. Both quantities start with unit weight. If a galaxy has a nearest neighbour (of the same target class) with a redshift failure ($w_{zf}$) or its redshift was not obtained because it was in a close pair ($w_{cp}$), we increase $w_{zf}$ or $w_{cp}$ by one. As found in \citet{ross2012}, the impact of this effect is very small for the CMASS sample, for this reason, we do not model the redshift failures in this study. For CMASS, additional weights are applied to account for the observed systematic relationships between the number density of observed galaxies and stellar density and seeing (weights $w_{star}$ and $w_{see}$, respectively).

\section{\textsc{BigMultiDark} simulation}
\label{sec:bigmd}

The \textsc{BigMDPL} is one of the \textsc{MultiDark}\footnote{ http://www.multidark.org/} $N$-body simulation described in \citet{klypin2014}. The \textsc{BigMDPL} was performed with GADGET-2 code \citep{springel2005}. This simulation was created in a box of 2.5 $h^{-1}$ Gpc on a side, with 3840$^3$ dark matter particles. The mass resolution is $2.4\times10^{10}$ $h^{-1}$ M$_{\odot}$. The initial conditions, based on initial Gaussian fluctuations, are generated with Zeldovich approximation at $z_{init} = 100$. The suite of \textsc{BigMultiDark} is constituted of four simulations with different sets of cosmological parameters. In this study, we adopt a flat $\Lambda$CDM model with the Planck cosmological parameters: $\Omega_m=0.307$, $\Omega_B=0.048$, $\Omega_\Lambda=0.693$, $\sigma_8=0.829$, $n_s=0.96$ and a dimensionless Hubble parameter $h=0.678$ \citep{klypin2014}. The simulation provides twenty redshift outputs (snapshots) within the redshift range $0.43<z<0.7$.

For the present analysis, we use the \textsc{RockStar} (Robust Overdensity Calculation using K-Space Topologically Adaptive Refinement) halo finder \citep{behroozi2013}. Spherical dark matter halos and subhalos are identified using an approach based on adaptive hierarchical refinement of friends-of-friends groups in six phase-space dimensions and one-time dimension. \textsc{RockStar} computes halo mass using spherical overdensities of a virial structure. Before calculating halo masses and circular velocities, the halo finder performs a procedure which removes unbound particles from the final mass of the halo. \textsc{RockStar} creates particle-based merger trees. The merger trees algorithm \citep{behroozi2013-2} was used to estimate the peak circular velocity over the history of the halo, $V_{peak}$, which we use to perform the abundance matching.

\section{Methodology: The Survey Generator code}
\label{sec:sugar}

We construct light-cone catalogues from the \textsc{BigMDPL} simulation which reproduce the clustering measured in the monopole of the Redshift-space correlation function from the BOSS CMASS DR12 sample. For this purpose, we developed the SUrvey GenerAtoR code (\textsc{SUGAR}) which implements the HAM technique to generate galaxy catalogues from a dark matter simulation. The code can apply the geometric features of the survey and selection effects, including stellar mass incompleteness and fiber collision effects. All the available outputs (snapshots) of the \textsc{BigMDPL} simulation are used, so that the light-cone has the proper evolution of the clustering.

\begin{figure*}
   \includegraphics[width=180mm]{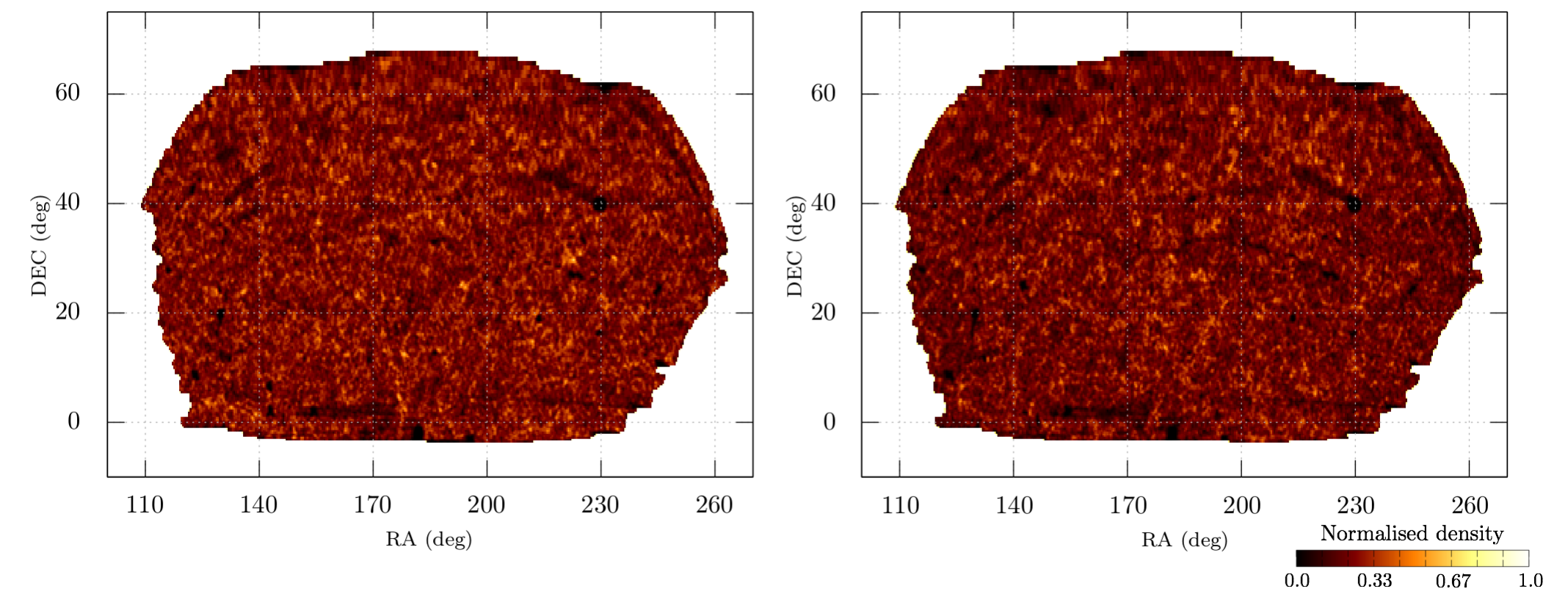}
     \caption{$Left$ $panel$: Sky area covered by the \textsc{BigMD-BOSS} light-cone. This region includes the BOSS CMASS DR12 geometry and veto masks. $Right$ $panel$: Sky area covered by the BOSS CMASS DR12 sample. Colours indicate the angular number density, which is normalised by the most dense pixel. Each pixel has an angular area of 1 deg$^2$. \textsc{BigMD}-BOSS light-cone uses the same mask as the BOSS CMASS DR12, including angular completeness and veto masks.}
  \label{fig:sky}
\end{figure*}

\begin{figure*}
  \includegraphics[width=170mm]{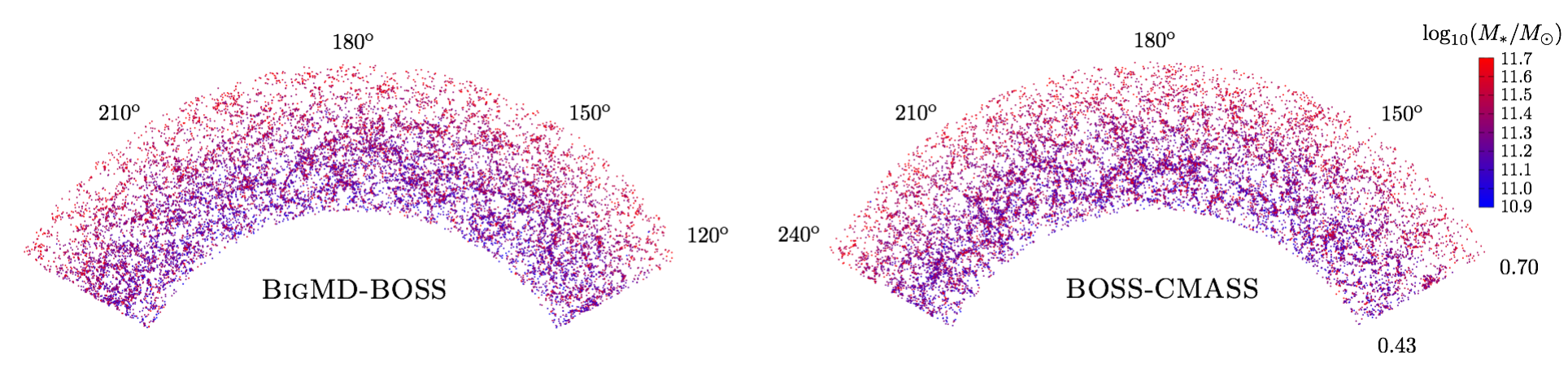}
  \caption{Pie plot of the \textsc{BigMD-BOSS} light-cone (left panel) and the BOSS CMASS DR12 data (right panel). Both figures were made with 2 deg of thickness (DEC coordinate).}
  \label{fig:piplot}
\end{figure*}

In the following subsections, we present the ingredients used to produce the \textsc{BigMD-BOSS} light-cone, which is showed in Figure \ref{fig:sky} and Figure \ref{fig:piplot}. We present the HAM method and the Stellar Mass Function (SMF) adopted in this work. The light-cone production, the fiber collision assignment and the modelling of the stellar mass incompleteness are also shown.

\subsection{Halo Abundance Matching procedure}
\label{sec:ham}

We use a HAM technique to populate dark matter halos with galaxies \citep[see e.g.,][]{nuza2013}. This physically motivated method produces mock galaxy catalogs that in the past  gave good representations  of large galaxy samples \citep[see for SDSS, e.g.,][]{trujillo2011,reddick2013}. The basic assumption of this method is that massive halos host massive galaxies. This allows one to generate a rank-ordered relation between dark matter halos and galaxies. However, observations show that this assignment cannot be a one-to-one relation \citep{shu2012}. In order to create a more realistic approach, it is necessary to include scatter in this matching. The HAM can relate galaxy luminosities or stellar mass from galaxies to a halo property. In this paper, we use the peak value of the circular velocity over the history of the halo ($V_{peak}$), which has advantages compared to the halo mass ($M_{halo}$). $M_{halo}$ is well-defined for host halos, but its definition becomes ambiguous for subhalos. The subhalo mass also depends on the halo finder used \citep{trujillo2011,reddick2013}. In addition to $M_{halo}$ and $V_{peak}$, HAM can be performed using other quantities such as the maximum circular velocity of the halo ($V_{max}$), the maximum circular velocity of the halo at time of accretion ($V_{acc}$) or the halo mass at time of accretion ($M_{acc}$). Other studies present the effect of the halo property in the HAM  \citep[e.g.,][]{reddick2013,guo2015b}

We adopt a modified version of the scatter proposed in \citet{nuza2013}. Our implementation of the abundance matching can be briefly summarised in the following steps:
\begin{enumerate}
\item For the dark matter halos, we define a scattered $V_{peak}$, which is used only to assign stellar mass to the halos. This scattered quantity is defined by:
  \begin{equation}
    \label{eq:scatter}
    V_{peak}^{scat}=(1+\mathcal{N}(0,\sigma_{\textsc{ham}}))V_{peak},
  \end{equation}
where $\mathcal{N}$ is a random number, produced from a Gaussian distribution with mean 0 and standard deviation $\sigma_{\textsc{ham}}(V_{peak}|M_*)$.
\item Sort the catalogue by $V_{peak}^{scat}$, starting from the object with the largest velocity and continuing down until reaching all the available objects. Use this catalogue to construct the cumulative number density of the halos as a function of  $V_{peak}^{scat}$.
\item Compute the cumulative number density of galaxies as a function of the stellar mass using the adopted SMF (see \ref{sec:stlmass}). 
\item Finally, construct a monotonic relation between the cumulative number density functions from step (ii) and (iii) such as
  \begin{equation}
    \label{eq:2}
    n_{gal}(>M_*^i)=n_{halo}(>V_{peak,i}^{scat}).
  \end{equation}
This relation implies that a halo with $V_{peak,i}^{scat}$ will contain a galaxy with stellar mass $M_*^i$.
\end{enumerate}
This assignment is monotonic between $V_{peak}^{scat}$ and $M_*$, but not between $V_{peak}$ and $M_*$. The relation of these two quantities is mediated by the scatter parameter, $\sigma_{\textsc{ham}}(V_{peak}|M_*)$. 

\subsection{Stellar Mass Function}
\label{sec:stlmass}

We employ the Portsmouth SED-fit DR12 stellar mass catalogue \citep{maraston2013} with the Kroupa initial mass function \citep{kroupa2001} to estimate the SMF. The CMASS large-scale structure (LSS) catalogue does not include the stellar mass information. For that reason, we matched the BOSS and the LEGACY stellar mass catalogues with the LSS BOSS CMASS catalogue. In order to identify a SDSS spectrum in the different catalogues, there are three numbers that determine each galaxy: \textsc{plate}, \textsc{mjd} and \textsc{fiberid}. We use these three quantities to match the stellar mass catalogues (LEGACY and BOSS) and the LSS BOSS CMASS catalogue. Once the stellar masses of the observed sample are assigned, we need to construct a SMF which describes the mass distribution.

\begin{figure}
  \includegraphics[width=84mm]{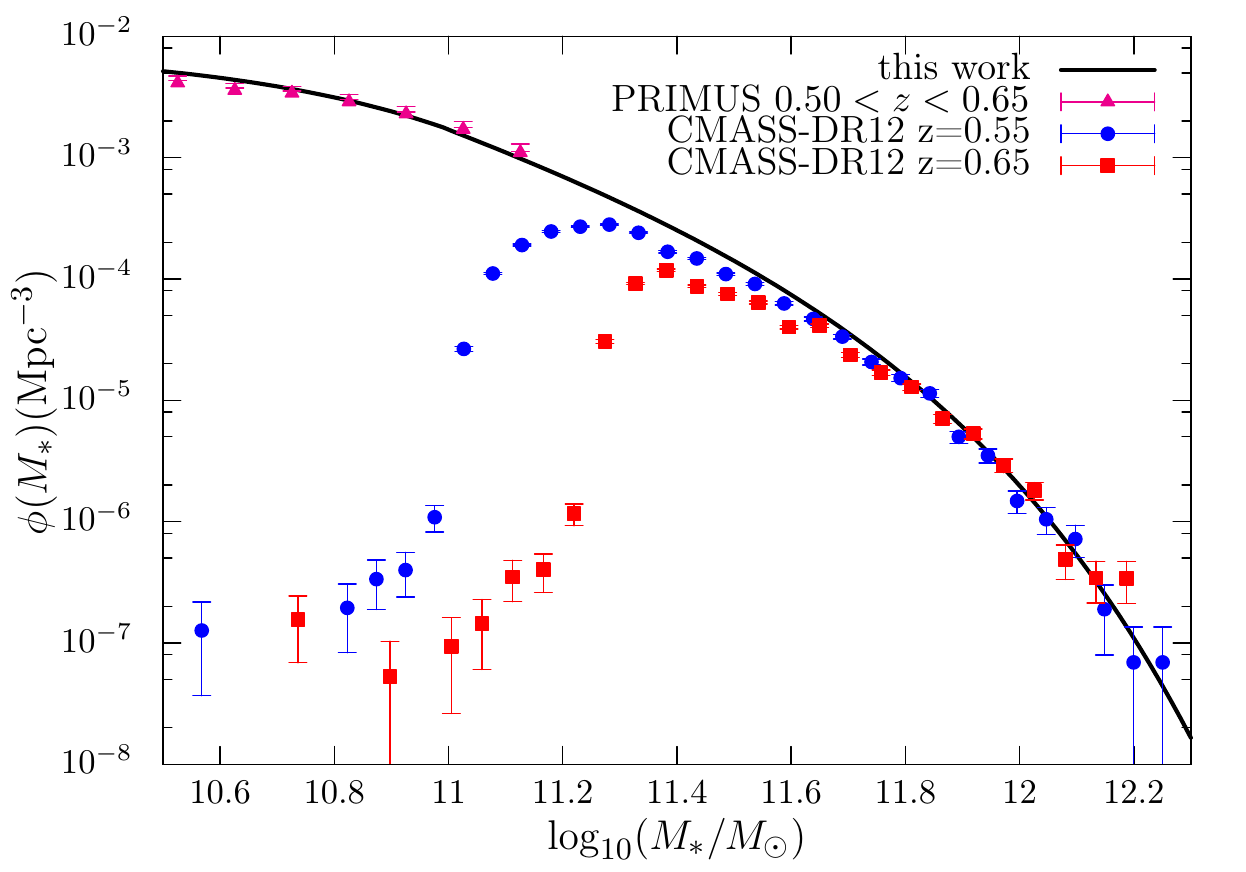}
  \caption{Stellar mass function from BOSS CMASS DR12 sample. Circles and squares show the stellar mass distribution for two redshift bins from the Portsmouth DR12 catalogue. Poissonian errors are included. The solid line shows the estimate of the SMF for this work, which is constructed combining the high-mass end of the BOSS sample and \citet{guo2010} for the low-mass range ($\log_{10}M*<11.0$). In order to compare with a complete sample in the redshift range 0.5 to 0.65, we include the PRIMUS SMF (Triangles) in the low-mass regime.}
  \label{fig:SMF}
\end{figure}

The Portsmouth DR12 catalogue has the SMF that is different from SMF of previous surveys \citep{maraston2013}. Figure \ref{fig:SMF} shows the mass distribution of the CMASS DR12 for two different redshift regions. A detailed study of the Portsmouth catalogues and other stellar mass catalogues was reported by \citet{maraston2013}. 

Due to the selection function in the BOSS data, we do not have the information on the shape of the stellar mass function at low masses. There are different ways of handling this problem. For example, \citet{leauthaud2015} use the stripe 82 massive galaxy catalogue to compute the SMF of the BOSS data. We use a different approach, for the high-mass end we use the Portsmouth stellar masses and we combine them with \citet{guo2010} results to describe the low-mass regime. Specifically, to compute the SMF for masses larger than $3.2\times10^{10}M_\odot$ (which is the mass range used in the CMASS sample). 
\begin{table}
  \centering
  \begin{tabular}{cccc}\hline
    Mass range&$\phi_*$&$\alpha$&$\log_{10}M_*$\\
    $[M_\odot]$ & $[$Mpc$^3\log_{10}M_\odot^{-1}]$ & & $[M_\odot]$\\ \hline
    $\log_{10} M_* \leq 11.00$&4.002$\times 10^{-3}$&-0.938&10.76\\
    $\log_{10} M_* > 11.00$ & 2.663$\times 10^{-3}$ &-2.447&11.42\\ \hline
  \end{tabular}
  \caption{Parameters of the double Press-Schechter SMF for this work.}
  \label{tab:param}
\end{table}

In order to construct the SMF, we select galaxies in the redshift range $0.55<z<0.65$, because this is the most complete range for the CMASS sample \citep[see][]{montero2014}. We combine the CMASS sample for masses larger than $2.5\times10^{11}M_\odot$ and the SMF from \citet{guo2010} for low masses. We fit both results using a double Press-Schechter mass function \citep{press1974} with the parameters given in Table \ref{tab:param}.

Figure \ref{fig:SMF} presents the SMF used in this work. We also add in Figure \ref{fig:SMF} the PRIMUS SMF  \citep{moustakas2013} in the redshift range $0.5<z<0.65$ with the purpose of comparing the low-mass range of our SMF with a complete sample in the same redshift and mass ranges. A detailed comparison of the Portsmouth catalogues and other stellar mass catalogues is presented in \citet{maraston2013}.

In our analysis we do not include redshift evolution of the stellar mass function. This approximation agrees with results of the PRIMUS survey \citep{moustakas2013}, which is a complete survey in the redshift range we study. \citet{moustakas2013}, show that there is only a small evolution of the stellar mass function in the CMASS redshift range.

\subsection{Production of Light-cones}

We implement a method to generate light-cones from snapshots of cosmological simulations. This method has been implemented previously \citep[see e.g.,][]{blaizot2005,kitzbichler2007}. The SUGAR code works with cubic boxes using positions and velocities of dark matter halos as inputs. We will now describe the procedure which we use to construct mocks for the CMASS sample.

\textsc{BigMD-BOSS} light-cones are constructed from the \textsc{BigMDPL} simulation which is large enough (2.5 $h^{-1}$ Gpc) to map the CMASS NGC. We use the periodic boundary conditions to maximise the use of the volume \citep{manera2013} but we do not reuse any region of the box.  So there are no duplicated structures in our light-cone.

\begin{figure}
  \includegraphics[width=84mm]{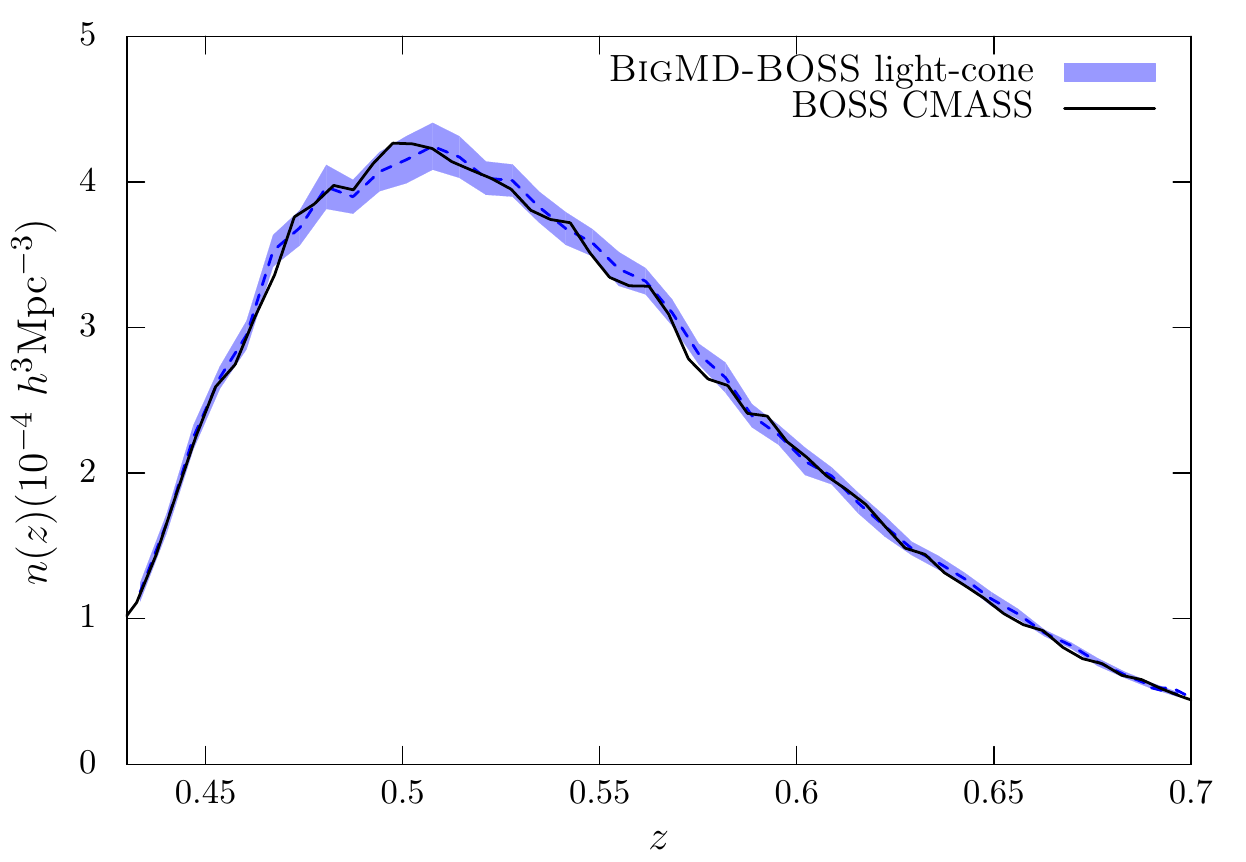}
  \caption{The comoving number density of BOSS CMASS DR12 NGC (black line) compared to the comoving number density of the \textsc{BigMD-BOSS} light-cone (Dashed line). Shaded area comes from 100 \textsc{md-patchy}  Mocks.}
  \label{fig:nbar}
\end{figure}

The first step in the construction of the light-cone is to locate the observer ($z=0$) and transform from comoving cartesian coordinates to equatorial coordinates (RA,DEC) and redshift. To include the effects of galaxy peculiar velocities in the redshift measurements, we transform the coordinates of the halos to Redshift-space using:
\begin{equation}
  \label{eq:rds}
   \mathbf{s} =\mathbf{r}_c + \frac{\mathbf{v}\cdot\hat{\mathbf{r}}}{aH(z_{real})},
\end{equation}
where $\mathbf{r}_c$ is the comoving distance in real space,  $\mathbf{v}$ is the velocity of the object with respect to Hubble flow, $\hat{\mathbf{r}}$ is the line of sight direction, $a$ is the scale factor and $H$ the Hubble constant at $z_{real}$, which is the redshift corresponding to $r_c$, and is computed from
\begin{equation}
\label{eq:comov}
r_c(z_{real})=\int\limits_0^{z_{real}}\frac{c dz}{H_0\sqrt{\Omega_m(1+z)^3+\Omega_\Lambda}},
\end{equation}
where $c$ is light speed and $H_0$ is the Hubble constant in s$^{-1}$ Mpc$^{-1}$ km. Using equation \eqref{eq:rds} and \eqref{eq:comov} it is possible to compute $s(z_{obs})$, where $z_{obs}$ is the observed redshift. The next step is to select objects from each snapshot to construct shells for the light-cone. Thus, an object with redshift $z_{obs}$, which comes from a snapshot at $z=z_i$, will be selected if $(z_i+z_{i-1})/2<z_{obs}\leq(z_i+z_{i+1})/2$. We repeat this process for all objects in snapshots between $z=0.43$ and $z=0.7$. We fix the number density in each shell following the radial selection function of the BOSS CMASS sample. Figure \ref{fig:nbar} shows the comparison between the radial selection function of the observed data and the one obtained on the \textsc{BigMD-BOSS} light-cone.

Finally, we apply the angular CMASS NGC mask to match the area of the observed sample. The angular completeness is taken into account by downsampling the regions where it is smaller than one. As was done in the BOSS CMASS catalogue, we select regions in the sky with completeness weight larger than 0.7. Due to the presence of random numbers in the selection process, the observed radial selection function can have variations of $\sim4\%$ . Figure \ref{fig:nbar} presents the standard deviation from 100 \textsc{md-patchy}  Mocks to examine the effect of different seed in the random generator.

Figure \ref{fig:sky} shows the angular distribution of the \textsc{BigMD-BOSS} light-cone.  In order to reproduce the angular distribution, we applied the BOSS CMASS DR12 NGC geometry, and, in addition, we applied veto mask to exclude exactly the same regions removed in the observed data. Figure \ref{fig:piplot} presents a 2D comparison of the spatial galaxy distribution between the \textsc{BigMD-BOSS} light-cone and the BOSS CMASS data.

\subsection{Stellar Mass Incompleteness}
\label{subsec:incom}

\begin{figure*}
  \includegraphics[width=168mm]{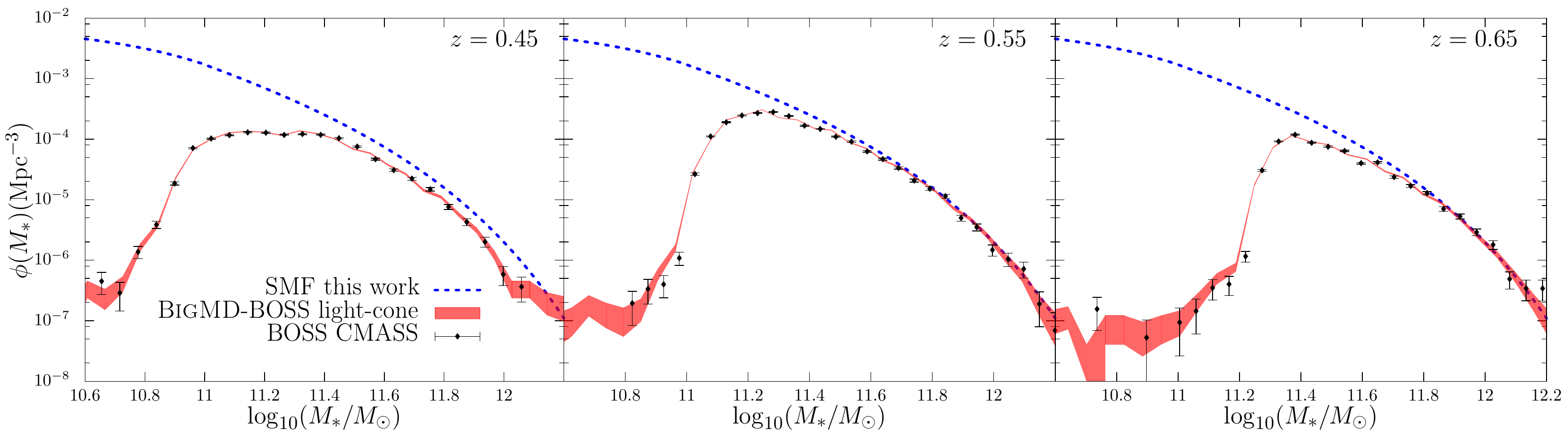}
  \caption{Incompleteness modelling for three different redshift bins. Shaded area shows the \textsc{BigMD-BOSS} light-cone, dots are the measurements from the CMASS Portsmouth catalogue. In both cases Poissonian errors are used. Dashed line represents the SMF adopted in this work. We select three bins as an example to show the results of the incompleteness modelling implemented in this work. Stellar mass distribution in the \textsc{BigMD}-BOSS light-cone are produced by downsampling galaxies from the SMF adopted. Left panel figure shows the incompleteness at low-redshift in the high-mass of the SMF.}
  \label{fig:incomp}
\end{figure*}

This paper focuses in the production of mocks which can describe the full CMASS DR12 sample. Instead of extracting a sub-sample which has better completeness in terms of stellar mass, we ``model'' the observed stellar mass incompleteness. This model not only accounts for the incompleteness at small masses (presented across the complete redshift range), but also incompleteness in the high-mass end, which is important for $z\lesssim0.45$. Figure \ref{fig:incomp} compares the results of our modelling in the \textsc{BigMD-BOSS} light-cone to the observed data for three different redshifts.

In order to reproduce the observed stellar mass distribution, we construct a continuous function by interpolation. Once the abundance matching is applied and galaxies are assigned to dark matter halos, we select galaxies by downsampling based on the observed stellar mass distribution. This process is repeated for 20 different redshifts (corresponding to the snapshots of the simulation). Then, in order to construct the observed stellar mass distribution corresponding to snapshot at $z=z_i$, a galaxy with redshift $z_{g}$ in the stellar mass catalogue will be selected if $(z_i+z_{i-1})/2<z_{g}\leq(z_i+z_{i+1})/2$. This model has an important impact on the scatter applied to the abundance matching. Since bias is as a function of stellar mass, incompleteness that varies as a function of stellar mass will affect the overall bias as well. This effect reduces the amplitude of the clustering, which implies that a smaller scatter is required to reproduce the signal of the observed clustering. If we ignore the incompleteness effect, we can still reproduce the clustering in the two point correlation function. However, this scatter is not the intrinsic one, and the final stellar mass distribution will not match the observed sample. \citet{favole2015b} show a similar model to reproduce the incompleteness of the ELG population from the BOSS sample.

Most galaxies in the CMASS sample are red galaxies. However, there is also a fraction of blue galaxies in the data. In addition, the blue sample is less complete than the red one \citep{montero2014}.The random downsampling of galaxies in the \textsc{BigMD}-BOSS light-cone does not distinguish between both populations, which can produce potential systematics due to the different completeness of both samples. In this study, we reproduce the observed stellar mass distribution by downsampling galaxies from a no-evolving SMF. However, SMF evolves with redshift, which can produce underestimation of the incompleteness for some ranges of stellar mass and overestimation for other ranges. 

\subsection{Fiber Collisions}

A feature of the BOSS fiber-fed spectrograph is that the finite size of the fiber housing makes impossible to place fibers within 62'' of each other in the same plate. This causes a number of galaxies to not have a fiber assigned and hence, there is no measurement of their redshift. We model the effect of fiber collisions as follows. A total of 5\% of the CMASS targets could not been observed due to the fiber collisions. These objects have an important effect at scales $\lesssim10$ $h^{-1}$ Mpc. In this paper, we model the fiber collision effect by adopting the method described in \citet{guo2012}.

The first step is to find the maximum number of galaxies that could be assigned fibers. This decollided sample ($D_1$) is a set of galaxies which are not angularly collided with other galaxies in this subsample. The second population ($D_2$) are the potentially collided galaxies. Each galaxy in this subsample is within the fiber collision scale of a galaxy in population 1. We must determine from the observed sample the fraction of collided galaxies ($D_2''$) in the $D_2$ group (i.e. $D_2''/D_2$) for sectors covered by different numbers of tiles.  Finally, we randomly select the fraction $D_2''/D_2$ to the $D_2$ galaxies in the mocks to be collided galaxies.

\begin{figure}
  \includegraphics[width=84mm]{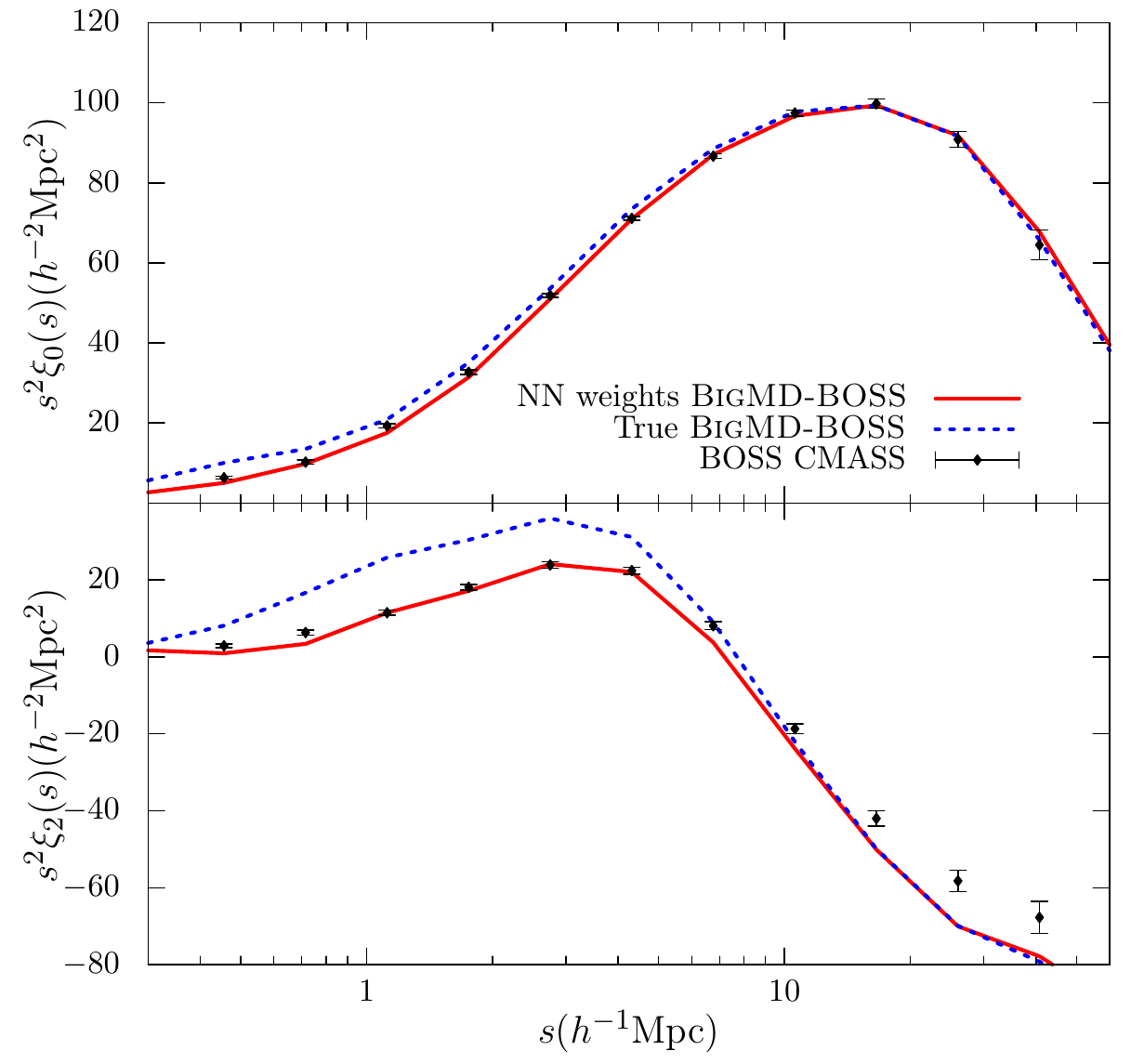}
  \caption{Monopole (top panel) and quadrupole (bottom panel) of the Redshift-space correlation function for the \textsc{BigMD-BOSS} light-cone before and after applying fiber collisions. Fiber collisions are corrected using nearest neighbour (NN) weights. The effects of the fiber collisions are stronger in the quadrupole, with important differences for scales $s\lesssim7$ $h^{-1}$ Mpc. The impact on the monopole is smaller. The fiber collision assignment is an approximative method which can introduce systematic effects. In order to avoid these effects, we select the range 2 $h^{-1}$ Mpc to 30 $h^{-1}$ Mpc to fit the monopole with the scatter parameter, $\sigma_{\textsc{ham}}(V_{peak}|M_*)$.}
  \label{fig:fiber}
\end{figure}

Figure \ref{fig:fiber} displays the impact of the fiber collisions in the correlation function in Redshift-space. The effect in the monopole becomes very important for scales smaller than 1 $h^{-1}$ Mpc. However, the quadrupole is more sensitive to this effect, with big impact for scales smaller to 10 $h^{-1}$ Mpc. The assignment of fiber collisions has an important impact on the fraction of satellites. Before fiber collisions the satellite fraction of the light-cone is 11.8\%, and after the assignment is equal to 10.5\%. This effect reduces the central-satellite pairs, which have a strong impact on the quadrupole.

Unlike \citet{guo2012}, we only use nearest neighbour weights for both samples. Our goal is to compare the results of the abundance matching with data, so that we implement the same fiber collision correction to our light-cone as observed data. 

\begin{figure}
  \includegraphics[width=84mm]{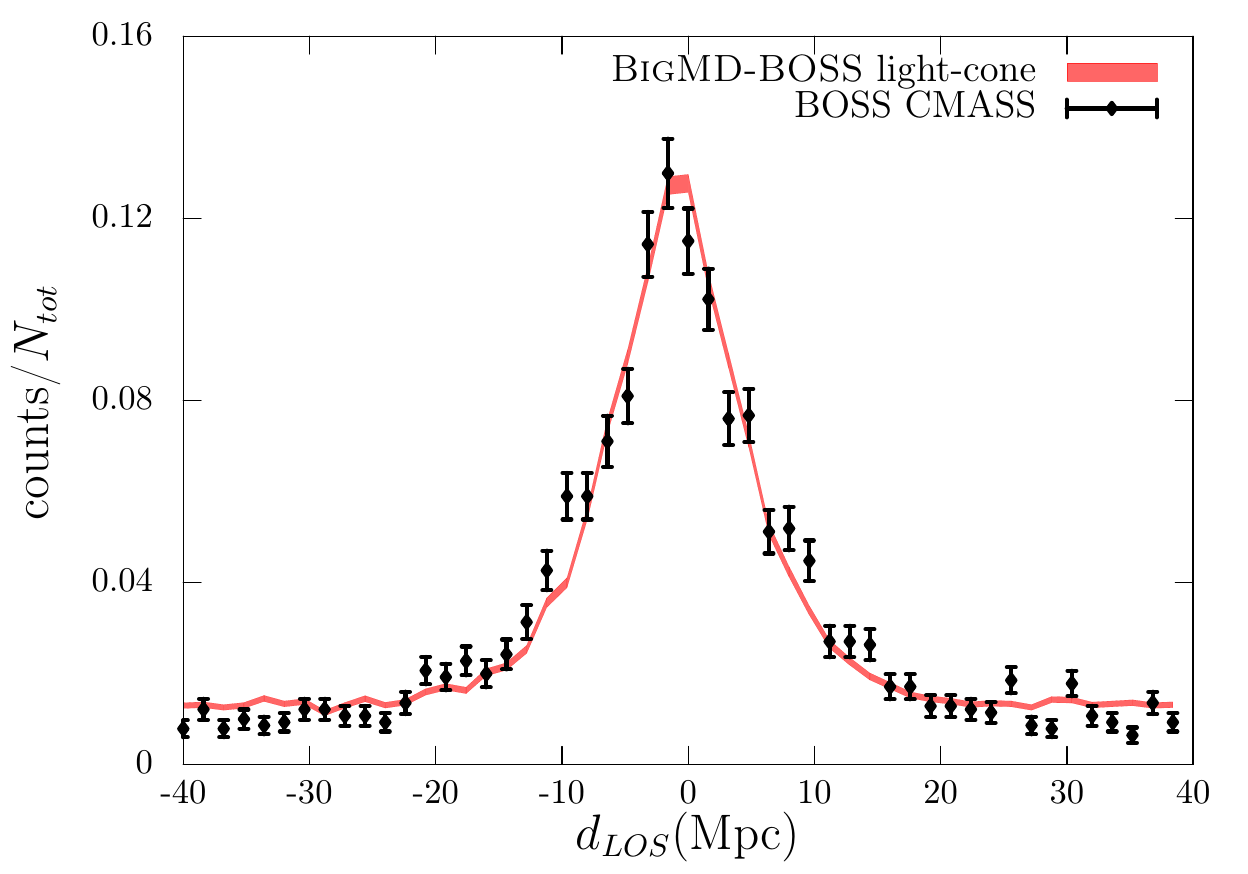}
  \caption{Line of sight displacement of a collided galaxy due to the fiber collision. The figure shows the number of counts per bin divided by the total number of collided galaxies. Uncertainties were computed using Poissonian errors.}
  \label{fig:dos}
\end{figure}

When nearest neighbour weights are applied, a collided galaxy will be ``moved'' from its original coordinates to the position of its nearest neighbour. Figure \ref{fig:dos} presents the line of sight displacement of those collided galaxies from their original positions. 

The displacement for the simulation shown in Figure \ref{fig:dos} is computed using the old and new positions of the collided galaxies. In CMASS data, the displacement is calculated using the overlapping tiled regions of the survey where the spectroscopic redshifts of both galaxies within the fiber collision angular scale are resolved. Figure \ref{fig:dos} demonstrates an excellent agreement between our model and the observed data, suggesting the combination between the clustering at small scales of the simulation and the fiber collision model used in the mock have a reasonable agreement with observations.

\section{Modelling BOSS CMASS clustering}
\label{sec:result}

The clustering signal in the abundance matching is determined by two quantities: the number density and the scatter in the $M_*-V_{peak}$ relation. The number density is fixed by the radial selection function of the observed sample. In order to find a scatter value that reproduces the clustering of the CMASS sample, we fit the monopole of the correlation function in Redshift-space. The following sections present the results of this monopole fitting, and the prediction of our model of the quadrupole in Redshift-space, projected correlation function, monopole in Fourier space and the Three-point correlation function.

\textsc{BigMD}-BOSS light-cone covers the same volume as CMASS sample between redshift $z$=0.43 and $z$=0.7. In order to have a good estimation of the uncertainties in our measurements we use 100 \textsc{md-patchy} mocks \citep[][companion paper]{kitaura2015}. These mocks are produced using five boxes at different redshifts that are created with the \textsc{patchy}-code \citep{kitaura2014}. This code matches the clustering of the galaxy catalogues for each redshift bin. The \textsc{md-patchy} mocks are based on the \textsc{BigMDPL} simulation, and they are produced with the same cosmology used in this work. To compute errors we use the square root of the diagonal terms of the covariance matrix defined as:
  \begin{equation}
    \label{eq:matrix}
    C_{ii}=\frac{1}{N-1}\sum\limits_{i=1}^N(X_i- \bar X)^2,
  \end{equation}
where $N$ is the number of mock catalogues and $X$ is the statistical quantity measured.

\subsection{Two point clustering: result from model and observations }

In order to compute the correlation function for our light-cone and the observed data, we use a Landy \& Szalay estimator \citep{landy1993}. The correlation function is defined by
\begin{equation}
  \label{eq:correst2D}
  \xi(r)=\frac{DD-2DR+RR}{RR}
\end{equation}
where $DD$, $DR$ and $RR$ represent the normalised data-data, data-random and random-random pair counts, respectively, for the distance range $[r-\Delta r/2, r+\Delta r/2]$.

In this paper we use random catalogues 20 times larger than the data catalogues. In order to estimate the projected correlation function and the multipoles of the correlation function we use the 2D correlation function, $\xi(r_p,\pi)$, where $s=\sqrt{r_p^2+\pi^2}$, $r_p$ is the perpendicular component to the line of sight and $\pi$ represent the parallel component. The correlation function of the \textsc{BigMD-BOSS} light-cone is computed using close pairs weights and FKP weights \citep{feldman1994},
\begin{equation}
  \label{eq:fkp}
  w_{\textsc{fkp}}=\frac{1}{1+n(z)P_{\textsc{fkp}}},
\end{equation}
where $n(z)$ is the number density at redshift $z$ and $P_{\textsc{fkp}}=20000$ $h^{-3}$ Mpc$^3$. We use the FKP weights to optimally weight regions with different number densities. In the case of the BOSS CMASS sample, we use the galaxy weights given in equation \ref{eq:weight} and in addition the FKP weights. The total weights for the data used in our analysis are $w_{tot}=w_{\textsc{fkp}}w_g$.

Note that $P_{\textsc{fkp}}$ is chosen to minimise the variance of power spectrum measurements. For the correlation function measurements, one should use the optimal weight from \citet{hamilton1993},
\begin{equation}
  \label{eq:wh}
  w_{\textsc{h}}=1/(1+n(z)J_w),  
\end{equation}
where
\begin{equation}
  \label{eq:jw}
  J_w=\int_0^r \xi(r) dV.   
\end{equation}
However, since we are fixing $w_{\textsc{fkp}}$ or $w_{\textsc{h}}$ to be a constant to simplify the computation, we expect that $w_{\textsc{h}}$ should be similar to $w_{\textsc{fkp}}$. In any case, the choice of optimal weight will not bias the measurements.

\subsubsection{Redshift-space correlation function}
\label{sec:rscf}

Previous works demonstrated the impact of the scatter in the clustering signal of a mock generated with the abundance matching \citep[e.g.,][]{reddick2013}. In this study, we search for a scatter parameter ($\sigma_{\textsc{ham}}(V_{peak}|M_*)$) which reproduces the monopole of the correlation function and provides the prediction for other quantities. The multipoles of the two-point correlation function, in Redshift-space, are defined by
\begin{equation}
  \label{eq:multipoles}
  \xi_l(s)=\frac{2l+1}{2}\int_{-1}^1\xi(r_p,\pi)P_l(\mu)d\mu
\end{equation}
where
\begin{equation}
  \label{eq:mu}
  \mu=\frac{\pi}{\sqrt{r_p^2+\pi^2}}
\end{equation}
and $P_l(\mu)$ is the Legendre Polynomial. We will present results for the monopole ($l=0$) and the quadrupole ($l=2$).

\begin{figure}
  \includegraphics[width=84mm]{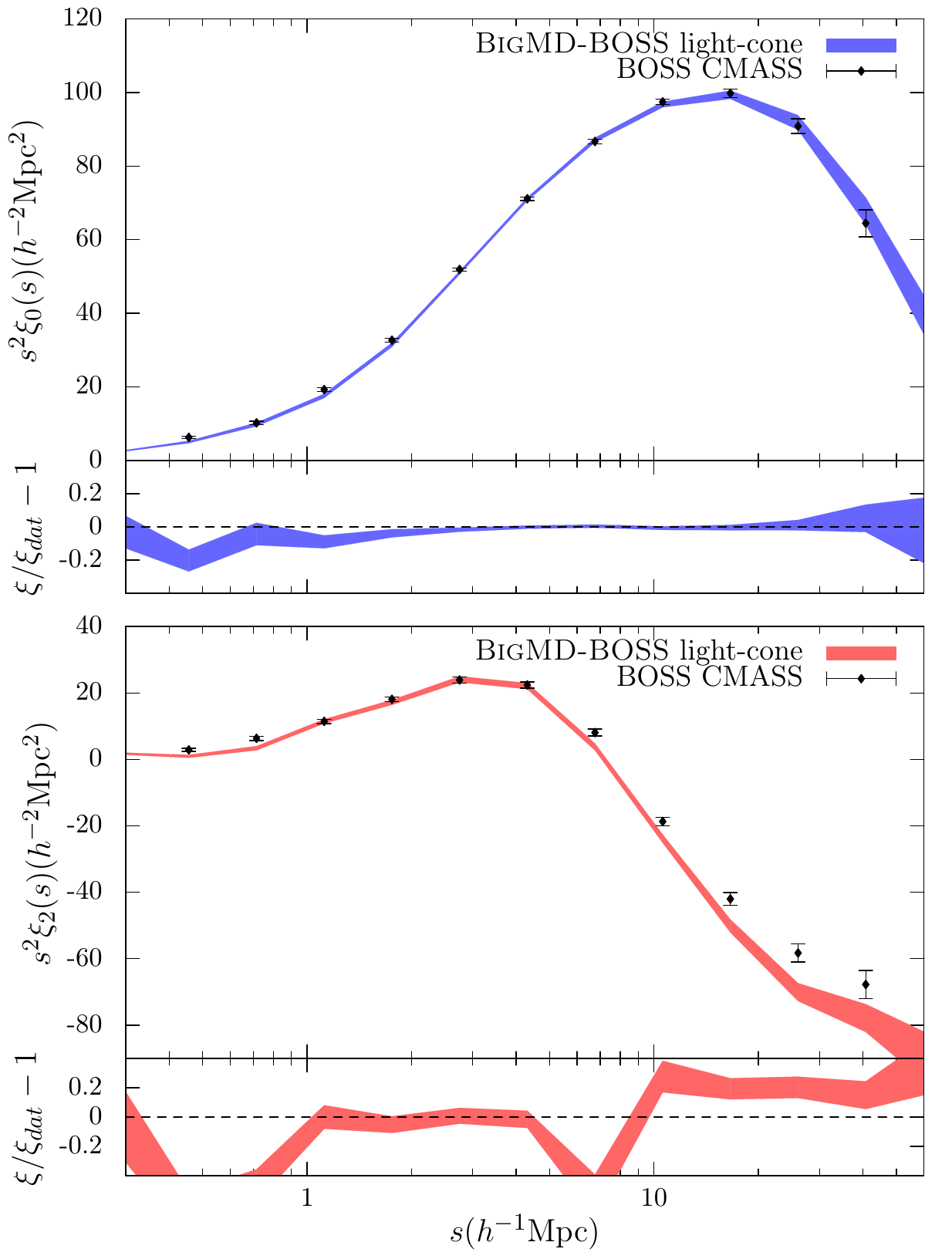}
  \caption{$Top$ $panel$: Monopole in Redshift-space from CMASS DR12 sample (black points). The shaded area represents the modelling of the monopole using the \textsc{BigMD-BOSS} light-cone. $Bottom$ $panel$: Quadrupole in Redshift-space from CMASS DR12 sample compared with the theoretical prediction from the \textsc{BigMD-BOSS} light-cone. Error bars were computed using \textsc{md-patchy}  mocks. Small panels show the ratio between the model and the observed data. Fitting of the monopole is performed between 2 $h^{-1}$ Mpc and 30 $h^{-1}$ Mpc. The observed monopole is in good agreement with our model for scales larger that 2 $h^{-1}$ Mpc. However, the quadrupole shows tensions with observations for scales $<1$ $h^{-1}$ Mpc and $5>$ $h^{-1}$ Mpc.}
  \label{fig:corr-log}
\end{figure}

\begin{figure}
  \includegraphics[width=84mm]{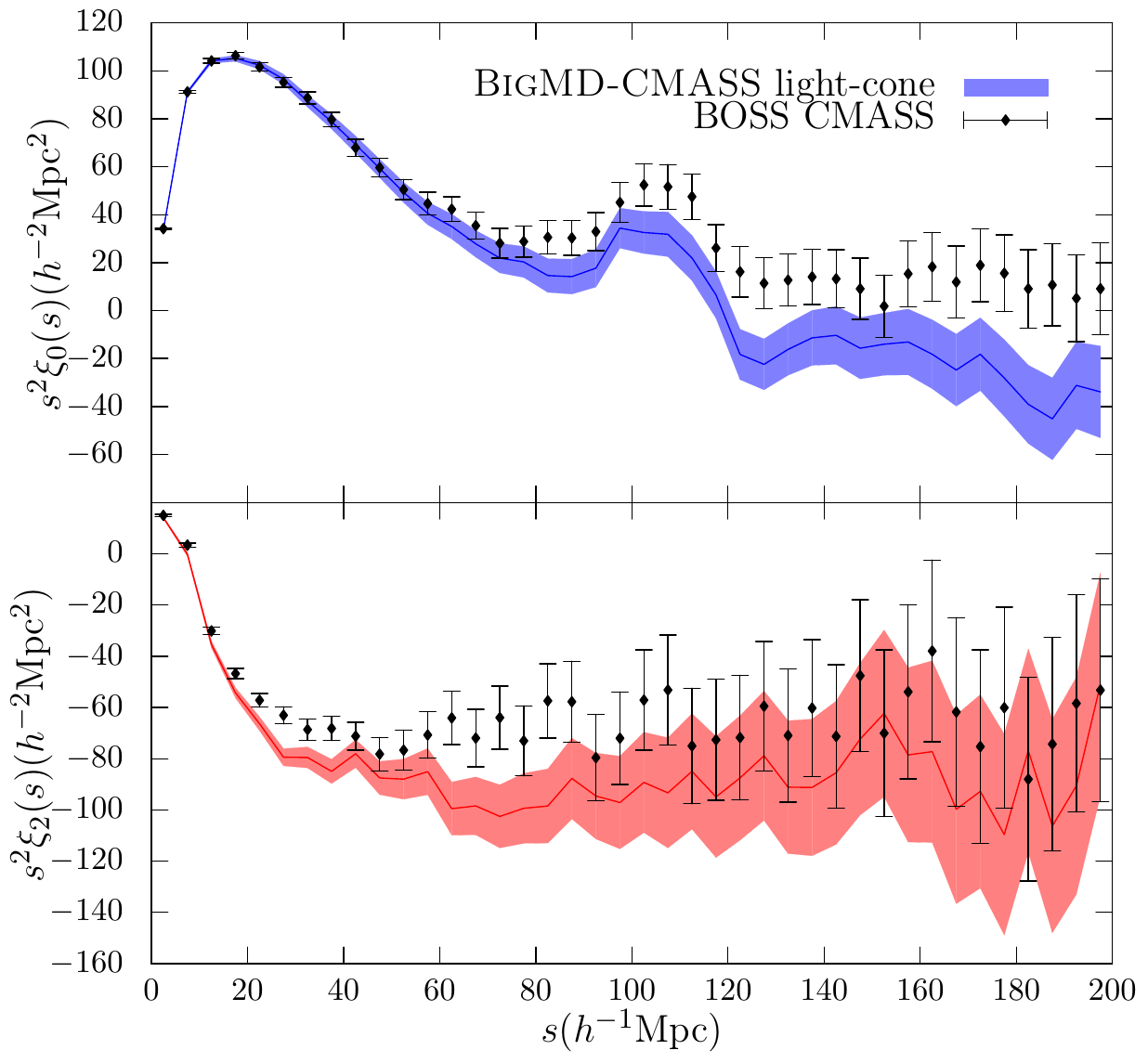}
  \caption{Monopole (top panel) and quadrupole (bottom panel) of the Redshift-space correlation function. The shaded areas are the model predictions for large scales using a single light-cone. Error bars were computed using \textsc{md-patchy}  mocks. Differences in the quadrupole are the same showed in the Figure \ref{fig:corr-log}. The monopole has a good agreement  up to 100 $h^{-1}$ Mpc. However, large scales present significant difference, but this can be due to the cosmic variance and remaining systematics in the data. These differences are within 2-$\sigma$ errors.}
  \label{fig:corr-lin}
\end{figure}

To find the best value, we fit the clustering using the monopole in the Redshift-space for the range 2 to 30 $h^{-1}$ Mpc. Top panel in Figure \ref{fig:corr-log} shows the results of the fitting compared to the CMASS DR12 data. Errors in Figure \ref{fig:corr-log} and in Figure \ref{fig:corr-lin} are computed using 100 \textsc{md-patchy}  mocks \citep[][companion paper]{kitaura2015}. The parameter that best reproduces the clustering in the monopole is $\sigma_{\textsc{ham}}(V_{peak}|M_*)=0.31$. This result is in agreement with previous works on abundance matching \citep{trujillo2011,nuza2013,reddick2013}.

The simulation provides a good agreement with data in the monopole for scales smaller than 50 $h^{-1}$ Mpc. However, the bottom panel in Figure \ref{fig:corr-log} shows a disagreement in the quadrupole for scales smaller than 0.7 $h^{-1}$ Mpc, that can be due to the method used to assign the fiber collisions in the \textsc{BigMD-BOSS} light-cone, for this reason, we do not analyse these scales. An additional disagreement is found at scales larger than 6 $h^{-1}$ Mpc, which will be commented in the last section of this work. \citet{nuza2013} use the \textsc{MultiDark} simulation with $\Omega_m=0.27$. Comparing their results for the monopole, we obtain a better agreement for scales larger than 10 $h^{-1}$ Mpc, mainly due to the difference in cosmologies used in this work. 

Figure \ref{fig:corr-lin} shows the prediction of the monopole and quadrupole for large scales compared to the observed data. Discrepancies for some values between the model and the data at scales larger than 60 $h^{-1}$ Mpc, could not be due only to the cosmic variance. Differences at the baryon acoustic oscillation (BAO) scales are of the order of 1 sigma errors while for large scales differences can be of the order of  2 or 3 sigmas. In Figure \ref{fig:corr-lin}, we can see that the BOSS CMASS correlation function at large scales is systematically shifted. This excess of power in the correlation function monopole could be due to the potential photometric calibration systematics which only affect very large scales. \citet{huterer2013} make a detailed study about the photometric calibration errors and their implication in the measurements of clustering and demonstrate that calibration uncertainties generically lead to large-scale power.

\subsubsection{Projected correlation function}

The projected correlation function is a quantity which is insensitive to the impact of the Redshift-space distortion and provides an approximation to the real space correlation function \citep{davis1983}. The projected correlation function is defined as the integral of the 2D correlation function, $\xi(r_p,\pi)$, over the line of sight:
\begin{equation}
  \label{eq:wpint}
  w_p(r_p)=2\int\limits_0^\infty \xi(r_p,\pi)\text{d}\pi.
\end{equation}

\begin{figure}
  \includegraphics[width=84mm]{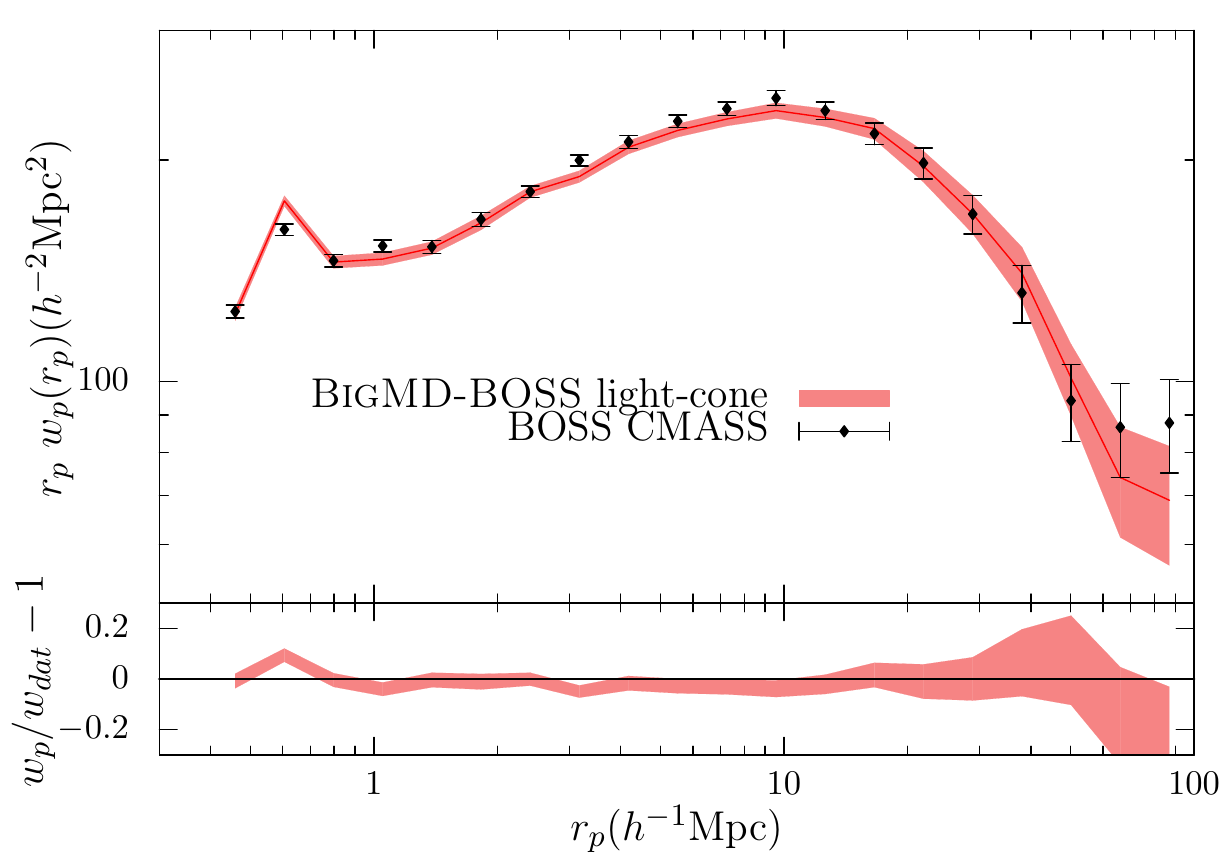}
  \caption{Projected correlation function prediction from the \textsc{BigMD-BOSS} light-cone (shaded region) compared to the BOSS CMASS sample. The width of the shaded area represents 1-$\sigma$ errors, computed using \textsc{md-patchy}  mocks. Our model reproduces the clustering for all relevant scales. Scales $<0.6$ $h^{-1}$ Mpc are dominated by fiber collision effects.}
  \label{fig:projected}
\end{figure}

In order to compute $w_p(r_p)$ from the discrete correlation function (equation \eqref{eq:correst2D}), we use the estimator:
\begin{equation}
  \label{eq:wp}
  w_p(r_p)=2\sum\limits_i^{\pi_{max}} \xi(r_p,\pi_i)\Delta\pi_i.
\end{equation}
We adopt a linear binning in the light of sight direction, $\Delta\pi_i=\Delta\pi=5$ $h^{-1}$ Mpc. We selected $\pi_{max}=100$ $h^{-1}$ Mpc. \citet{nuza2013} find convergence of the projected correlation for this scale. Figure \ref{fig:projected} shows the results found for the \textsc{BigMD-BOSS} light-cone compared to the CMASS data. Error bars were computed using 100 \textsc{md-patchy}  mocks.

Figure \ref{fig:projected} reveals a discrepancy  at scales $\approx 3$ $h^{-1}$ Mpc. However, results are in agreement at 2-$\sigma$ level, so we can consider the data consistent with the prediction of our model. Scales below 0.5 $h^{-1}$ Mpc are dominated by fiber collision. Due to this effect, the clustering declines rapidly.

\subsubsection{Fourier space}

The power spectra for the BOSS CMASS sample with nearest angular neighbour upweighted weights and the \textsc{BigMDPL} are computed using the \citet{feldman1994} power spectrum estimator modified to account for the systematic weights of the galaxies. In BOSS CMASS, each galaxy is assigned a systematics weight (equation \eqref{eq:weight}), which is accounted for in the estimator. For the \textsc{BigMD-BOSS} light-cone, we set  $w_g = w_{cp}$, for the power spectrum using nearest neighbour upweighted fiber collisions weights, and $w_g = 1$  for the true power spectrum. 
\begin{figure}
  \includegraphics[width=84mm]{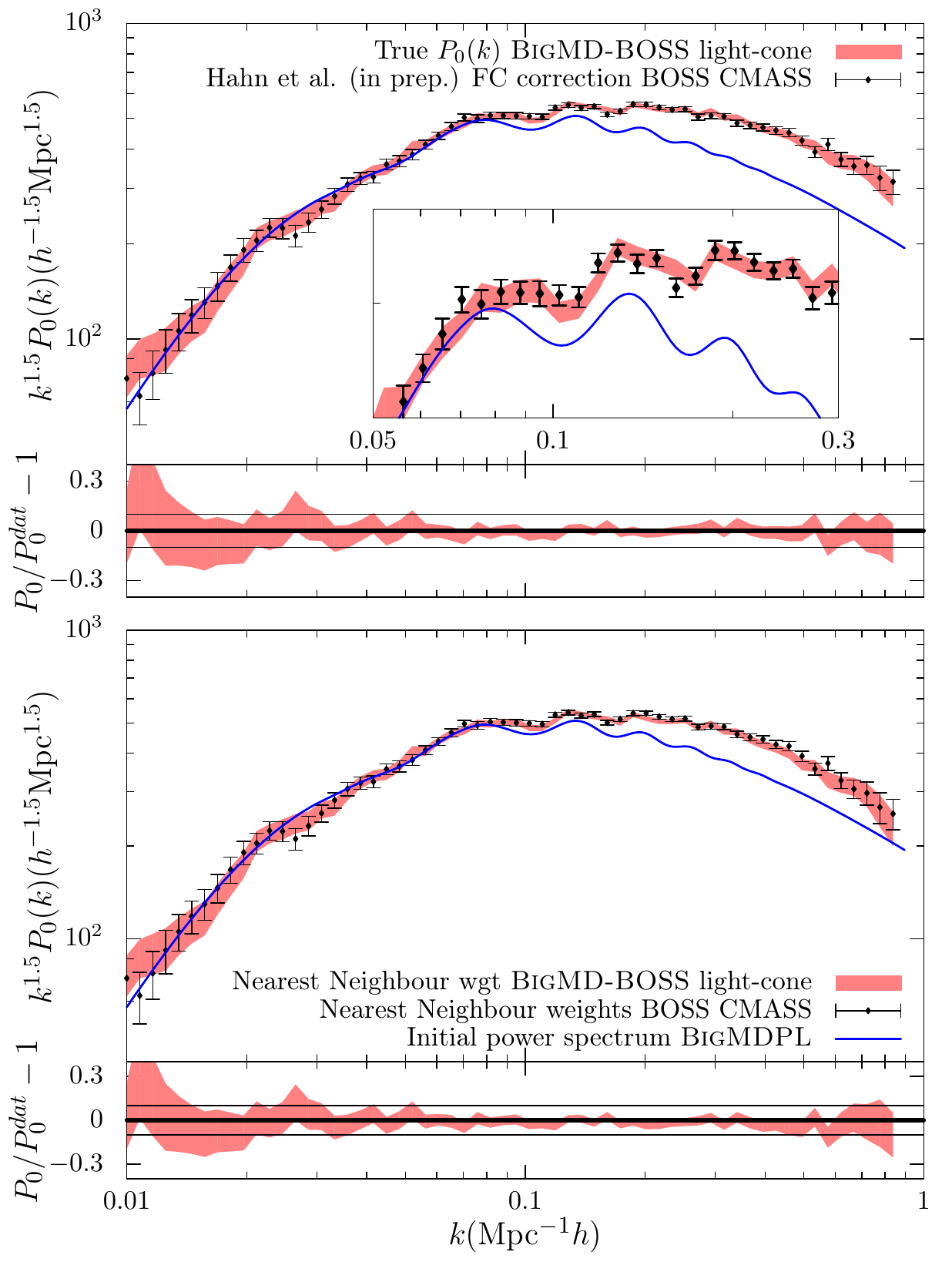}
  \caption{Monopole of power spectrum from the \textsc{BigMD-BOSS} light-cone and the CMASS DR12 sample. $Top$ $Panel$: The true power spectrum for our light-cone compared to the CMASS DR12 data corrected by fiber collisions using Hahn et al. (in prep.) method. Solid curve shows the initial matter power spectra of the \textsc{BigMDPL} simulation scaled to match the amplitude of fluctuations at long waves. A remarkable agreement between the data and the model is found for scales $k$ $\lesssim1$ $h$ Mpc$^{-1}$. $Bottom$ $Panel$: The comparison between simulation and observed data using nearest neighbour weights ($w_{cp}$). In addition to $w_{cp}$, observed measurements include systematics weights: $w_{star}$, $w_{zf}$ and $w_{see}$. The agreement between the data and the model, in both panels, shows the good performance of the fiber collisions assignment in the light-cone. In bottom subpanels, dashed lines represent an accuracy level of 10\%}
  \label{fig:pk}
\end{figure}

The power spectrum for the BOSS CMASS sample is computed using the method described in Hahn et al. (in prep.) in order to correct the effects of fiber collisions on smaller scales. The fiber collision correction method reconstructs the clustering of fiber-collided pairs by modelling the distribution of the line-of-sight displacements between them using pairs with measured redshifts. In addition, the method corrects fiber collisions in the shot-noise correction term of the power spectrum estimator. In simulated mock catalogues, the correction method successfully reproduces the true power spectrum with residuals $\lesssim1$\% at $k\sim0.3$ $h$ Mpc$^{-1}$ and  $<10$\% at $k\sim 0.9$ $h$ Mpc$^{-1}$. Top panel of Figure \ref{fig:pk} compares the fiber collision and systematics corrected BOSS CMASS power spectrum to the true power spectrum of \textsc{BigMD}-BOSS light-cone, showing remarkably good agreement between data and model. Figures \ref{fig:corr-log} and \ref{fig:pk} confirm that the standard HAM is accurate in the modelling of the clustering not only at large scales, but also in the one halo term.

Monopoles from our model and the BOSS CMASS data using fiber collision weights are shown in the bottom panel of Figure \ref{fig:pk}. Both power spectra agree for $k$ smaller than 1 $h$ Mpc$^{-1}$. The \textsc{BigMD-BOSS} light-cone and the observed data have a remarkably good agreement in the BAO region (inset panel Figure \ref{fig:pk}), which is not seen in the correlation function (Figure \ref{fig:corr-lin}). This difference can be due to remaining systematics that have a bigger impact on the correlation function than in the power spectrum. The agreement between our model and the observed data, for the true power spectrum and the nearest neighbour corrected power spectrum, demonstrates that the method used to assign fiber collisions in the \textsc{BigMD-BOSS} light-cone is a good approach to simulate this effect.

As we discussed in Section \ref{sec:rscf}, the disagreement between the model and the data in the correlation function monopole could be due to potential photometric calibration systematics. The effect on the power spectrum will be limited to very small $k$, so that it has less impact on the BAO scales. However, this excess of power does not have impact on BAO measurements from correlation functions when we marginalise the overall shape \citep[see][Ross et al. in prep.]{chuang2013}.

\subsection{Three-point correlation function}

We are also interested in comparing the prediction of the Three-point correlation function using the HAM on the \textsc{BigMDPL} simulation with the observed data. The 3PCF provides a description of the probability of finding three objects in three different volumes. In the same manner as the 2PCF, the 3PCF is defined as:
\begin{equation}
  \label{eq:3pcf}
  \zeta(r_{12},r_{23},r_{31})=\langle\delta(r_1)\delta(r_2)\delta(r_3)\rangle,
\end{equation}
where $\delta(r)$ is the dimensionless overdensity at the position $r$ and $r_{ij} = r_i - r_j$. We use the Szapudi \& Szalay estimator \citep{szapudi1998},
\begin{equation}
  \label{eq:esti3pcf}
  \zeta=\frac{DDD-3DDR+3DRR-RRR}{RRR}.
\end{equation}

\begin{figure}
  \includegraphics[width=84mm]{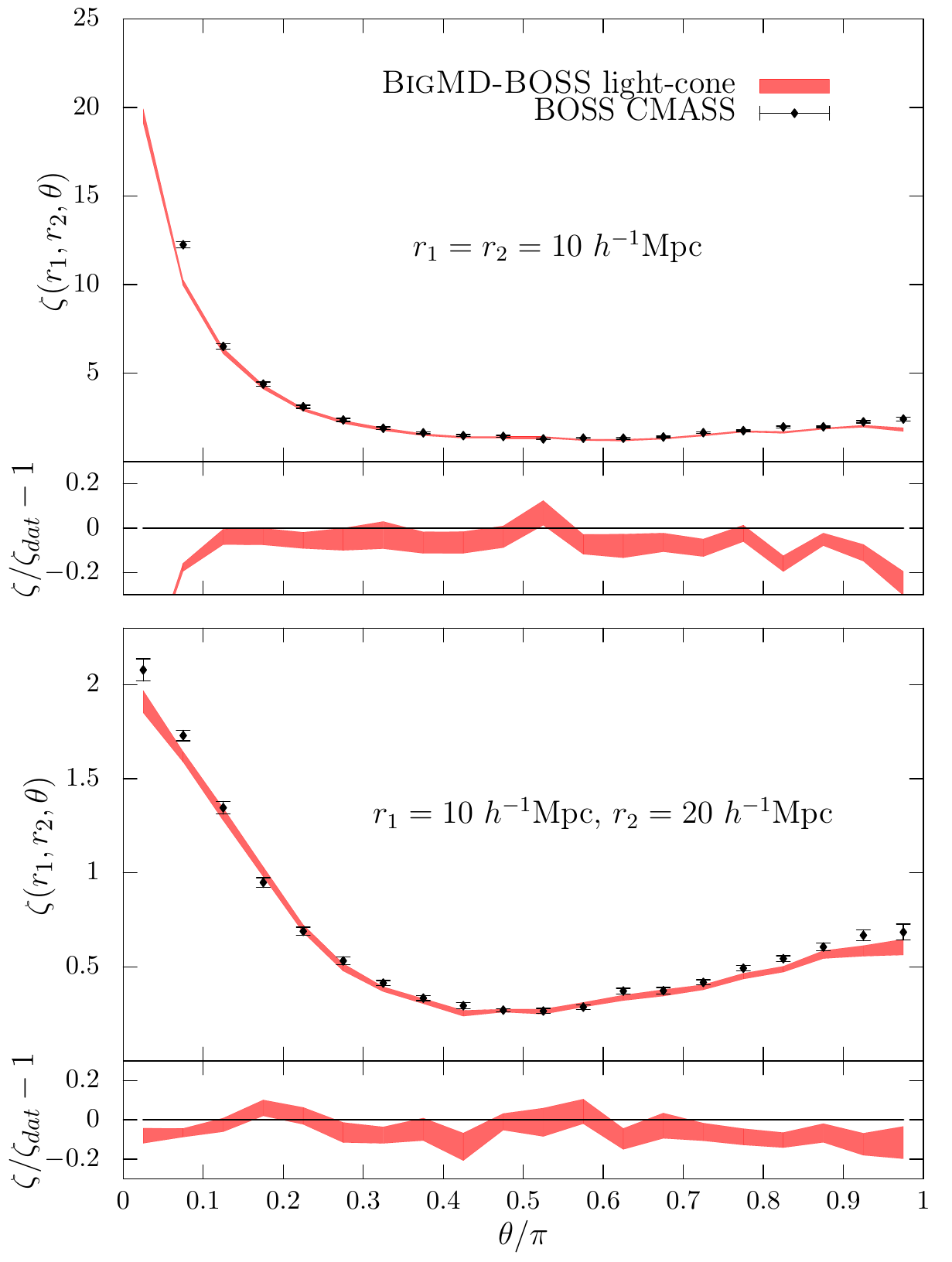}
  \caption{$Top$ $panel$: BOSS CMASS DR12 Three-point correlation function compared with the model prediction of this work. Shaded area shows 1-$\sigma$ uncertainties, with limits $r_1=10$ $h^{-1}$ Mpc and $r_2=20$ $h^{-1}$ Mpc. $Bottom$ $panel$: Three-point correlation function for limits $r_1=r_2=10$ $h^{-1}$ Mpc. The \textsc{BigMD}-BOSS light-cone can reproduce almost all scales between 2-$\sigma$ errors.}
  \label{fig:3pcf}
\end{figure}

Figure \ref{fig:3pcf} displays our prediction compared with the BOSS CMASS data. We see the results for two kinds of triangles: $r_1=r_2=10$ $h^{-1}$ Mpc and $r_1=10$ $h^{-1}$ Mpc, $r_2=20$ $h^{-1}$ Mpc, where $\theta$ is the angle between $r_1$ and $r_2$.

A good agreement in the shape of the 3PCF is seen in Figure \ref{fig:3pcf} between our prediction and the data. Most of the points are in agreement within 2-$\sigma$ errors for both configurations represented in Figure \ref{fig:3pcf}. However, the \textsc{BigMD}-BOSS light-cone is underestimating the 3PCF for $\theta\sim0$ and $\theta\sim\pi$. \citet{guo2015c} find similar discrepancies for those scales, which can be produced by velocity effects and can be corrected including a velocity bias. Therefore, the disagreement in the Three-point correlation function and in the quadrupole of the correlation function can be caused by the same kind of effects.

\subsection{Stellar to halo mass relation}

\begin{figure}
  \includegraphics[width=84mm]{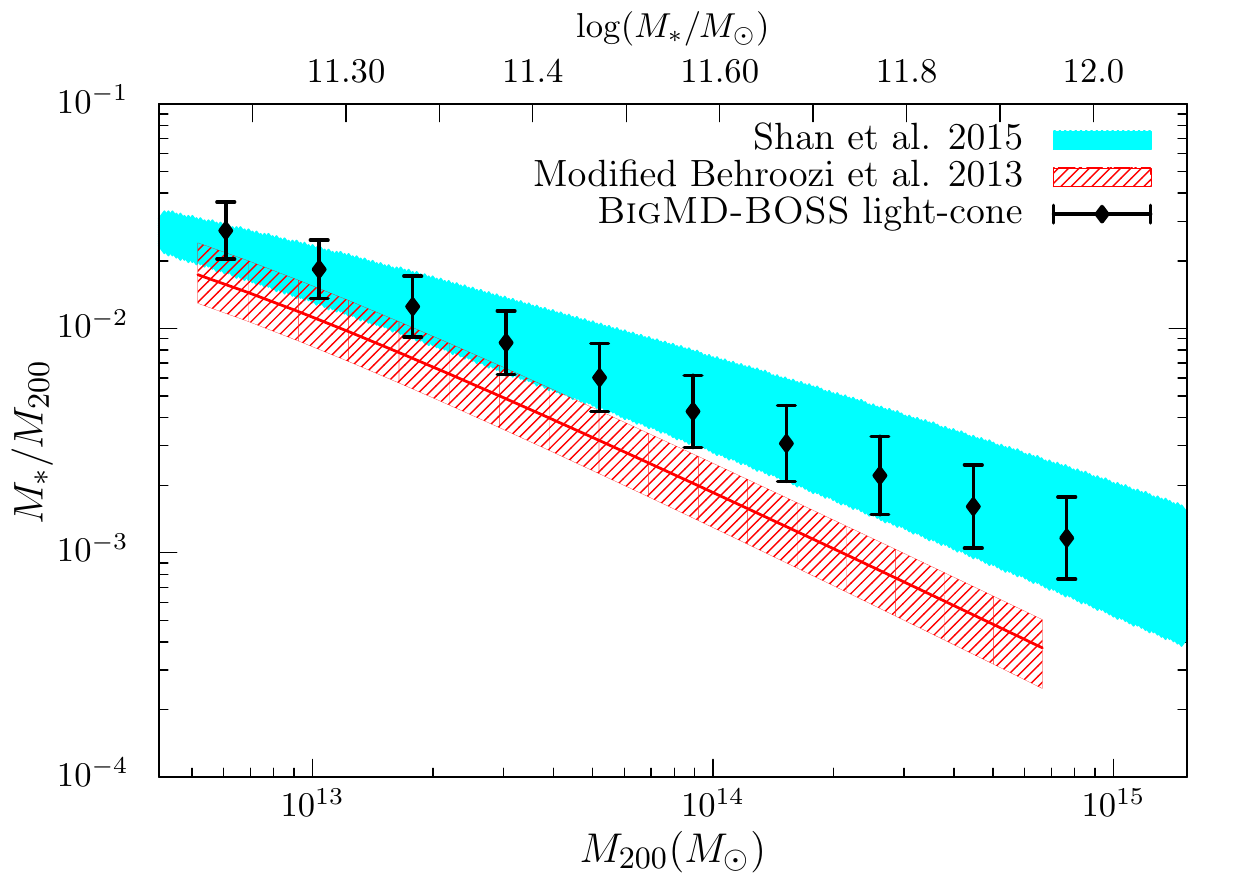}
  \caption{Stellar-to-halo mass ratio. The shaded blue area represents the best fit of the stellar to halo mass relation measured using weak lensing in the CFHT Stripe 82 Survey (Shan et al. 2015). The red area represent previous HAM result from Behroozi et al. 2013c. The analysis in \citet{behroozi2013-3} was modified using the Planck cosmology parameters and changing the definition of the halo mass.  Black dots are the prediction from the HAM - \textsc{BigMD-BOSS} light-cone. Differences between our model and Behroozi et al. 2013c. are mainly due to the SMF adopted in both works. Scatter between $M_{200}$ and $M_*$ is similar between the data and our model. We adopted constant scatter while observed data suggests a dependency of the scatter with the stellar mass. }
  \label{fig:SHMR}
\end{figure}

The Stellar to Halo Mass Ratio (SHMR) is an important quantity to evaluate if the simulated light-cone is providing a realistic halo occupation. In this way, we use results from weak lensing, which is one of the most powerful mechanisms to know the observational SHMR. Figure \ref{fig:SHMR} shows the SHMR predicted by the \textsc{BigMD-BOSS} light-cone and measurements in the CFHT Stripe 82 Survey \citep{shan2015}. In order to ensure the convergence of the halos in our prediction, we select halos with masses larger than 5.2$\times10^{12}M_\odot$. This limit is 150 dark matter particles which give convergence for subhalos \citep{klypin2015}. 

Predictions of the abundance matching are in agreements with the weak lensing data. In Figure \ref{fig:SHMR}, shaded blue area shows the intrinsic scatter measured. The dependency between scatter and stellar mass is clear. It is also shown in the abundance matching \citep[e.g.,][]{trujillo2011,reddick2013}. However, our HAM model uses a constant scatter to reproduce the clustering. This approximation can generate the disagreement in the scatter between data and mock. The red area in Figure \ref{fig:SHMR} indicates the results from \citet{behroozi2013-3}. We modify \citet{behroozi2013-3} in order to use the same definition of halo mass and implement the Planck cosmology in the analysis. The SMF assumptions can be one of the origins for the disagreement between both predictions. While we use the BOSS DR12 stellar mass catalogues to estimate the SMF, \citet{behroozi2013-3} use the PRIMUS SMF \citep{moustakas2013}. The difference in how the stellar mass catalogues handle profile fitting produce a variation in the high-mass end of both SMF. This effect causes important difference at large stellar mass between both predictions. 

\citet{shankar2014} present the stellar to halo mass relation assuming different mass functions and compare their results with recent models. They find differences between \citet{behroozi2013-3} and \citet{maraston2013} similar to the one shown in our Figure \ref{fig:SHMR}. \citet{shankar2014} also find that an intrinsic scatter in stellar mass at fixed halo mass of 0.15 dex is needed to reproduce the BOSS clustering. This result is in agreement with our model, which predicts an intrinsic scatter in stellar mass of 0.14 dex at a fixed halo mass. 

\subsection{Bias prediction}

Using the HAM-\textsc{BigMD-BOSS} light-cone and its corresponding dark matter light-cone we can estimate the real-space bias, $b(r)$, solving the equation \citep{kaiser1987,hamilton1992}
\begin{equation}
  \label{eq:bias}
  \xi(s)=\bigg(1+\frac{2}{3}\beta+\frac{1}{5}\beta^2\bigg)b(r)^2\xi_{DM}(r),
\end{equation}
where $\beta\approx f/b$ is the Redshift-space parameter and $f(z=0.55)=0.77$ (Planck cosmology).

\begin{figure}
  \includegraphics[width=84mm]{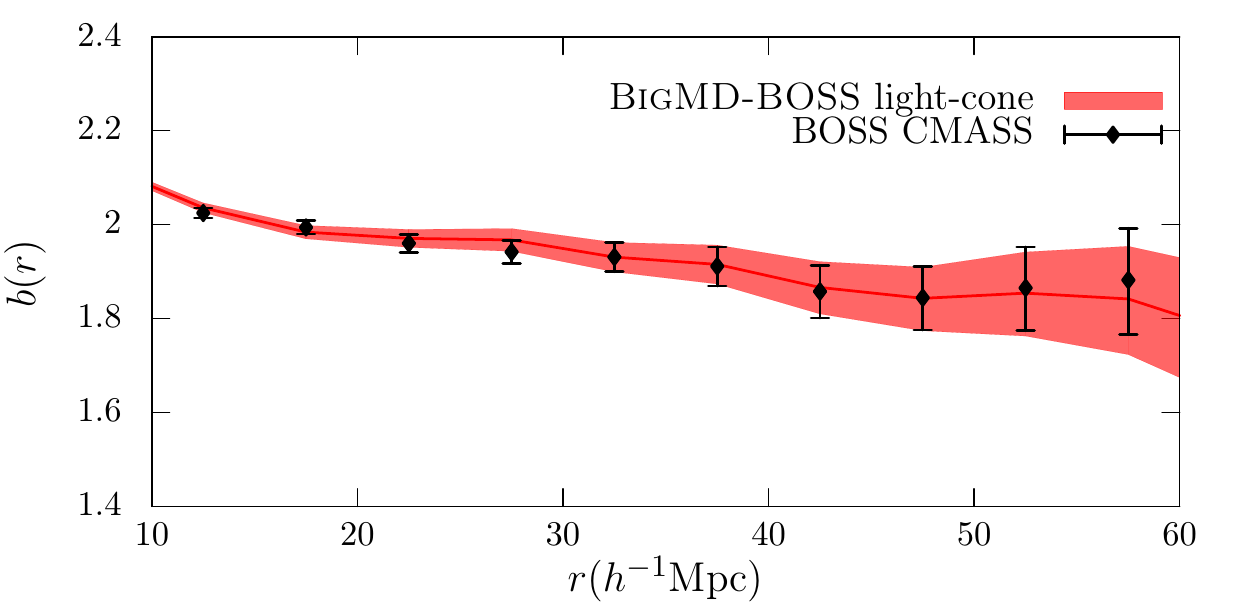}
  \caption{Scale-dependent galaxy bias from the model presented in this work. We measure the bias with respect to the correlation function of dark matter in the \textsc{BigMDPL} light-cone for the data and the model. There is an excellent agreement between the CMASS observations and the predictions of the HAM-\textsc{BigMD-BOSS} light-cone.}
  \label{fig:bias}
\end{figure}

Figure \ref{fig:bias} shows the linear bias, which is in agreement with previous papers that reproduced the CMASS clustering \citep[see][]{nuza2013}. For the data and the model, we use the dark matter correlation function from the \textsc{BigMD} simulation. For the scales shown, the scale-dependent bias factor is in the range 1.8-2. We use the \textsc{BigMD} dark matter light-cone to estimate the relative bias of the CMASS sample to this catalogue.

\section{Discussion}
\label{sec:discussion}

The \textsc{BigMD}-BOSS light-cone is designed to reproduce the full BOSS CMASS sample between redshift 0.43 to 0.7, including observational effects. In order to recover the information at small scales, similar papers \citep[e.g.,][]{guo2014,guo2015,nuza2013} correct the observed data by fiber collision \citep[see][Hahn et al. in prep.]{guo2012}. In this work, we assign fiber collisions to galaxies in the light-cone and we use nearest neighbour weights in the data and in the model. Our model can be useful to test methods that recover the clustering in the fiber collision region \citep{guo2012} or in the production of mocks for covariance matrices \citep[][companion paper]{kitaura2015}. The fiber collision assignment adopted in this work can reproduce in a good way this observational effect (Figure \ref{fig:pk}). However, this approach can introduce small systematics that we don't include in our modelling.

\citet{white2011} model the full CMASS clustering. They find a good fit of the HOD parameters to reproduce the observed data. However, they cannot describe the small scales because they only include close pair weights in the data measurements, which cannot recover the small scale clustering \citep{guo2012}. \citet{nuza2013} also reproduce with a good agreement the CMASS data using a standard HAM model, they correct by fiber collision using the method explained in \citet{guo2012}. Our paper continues the work presented in \citet{nuza2013}, including light-cone effects, redshift evolution, radial selection function, etc. All these papers can reproduce the clustering of the full CMASS sample.
\begin{figure*}
  \includegraphics[width=168mm]{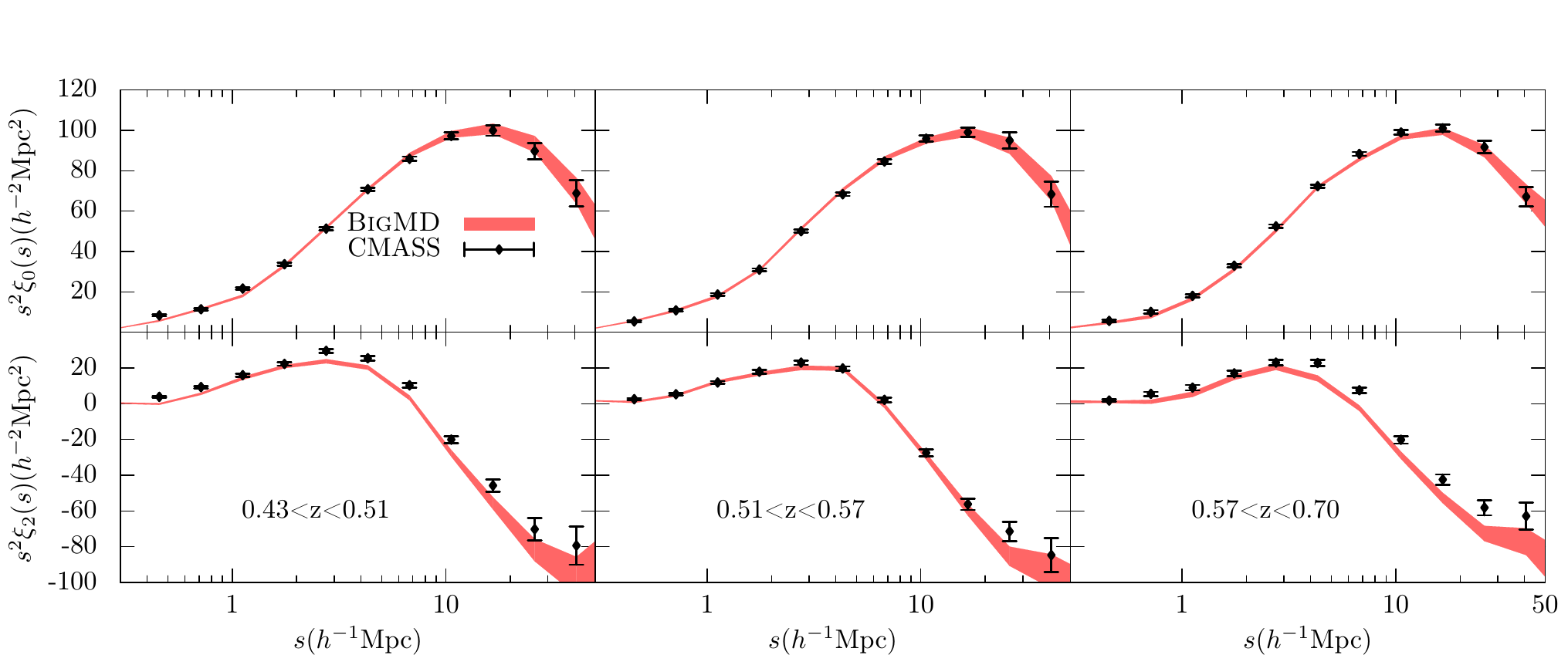}
  \caption{Monopole and quadrupole of the Redshift-space correlation function of the CMASS DR12 sample compared to the HAM-\textsc{BigMD-BOSS} light-cone for three redshift bins. The monopole is fitted for all redshift ranges. The middle bin is the most complete range in the CMASS sample, and also the best reproduced quadrupole.We perform a HAM with three different scatter parameters to fit each of the redshift bins. Differences at low and high redshift can be due to target selection effects we do not include in this study. Another source of discrepancy can be the relation between the scatter and the more massive galaxies (Saito et al. 2015).}
  \label{fig:quad-reds}
\end{figure*}

Recent papers show tensions between models and observed data when a most careful selection is done. \citet{guo2015} study a volume-limited luminous red galaxy sample in the redshift range of 0.48 < z < 0.55 of the CMASS sample. They need a galaxy velocity bias to describe the clustering of the most massive galaxies ($\sim10^{13}-10^{14}$ $h^{-1}$ M$_{\odot}$) using HOD. \citet{saito2015} show an extension of the HAM to describe the colour dependency of the clustering for the CMASS sample. \citet{guo2015b} present a comparison between HOD and HAM models, they also modify the standard HAM model in order to reproduce clustering at different luminosity cuts. \citet{favole2015} present a study of the blue population properties compared to the red galaxies. They present a modified HOD which allows to include both samples in the same mock catalogue. The clustering dependency on stellar mass (luminosity) is not implemented in our model and we do not distinguish between blue and red galaxies. Our implementation of the HAM and stellar mass incompleteness is capable of reproducing the full CMASS sample, including a big amount of data in our analysis. \citet{zu2015} present a modified HOD in order to include the stellar mass incompleteness (iHOD). This model combines galaxy cluster and galaxy--galaxy lensing and allows to increase $\sim80\%$ the number of modelled galaxies than the traditional HOD models.

We find the largest discrepancy between our model and the data in the quadrupole measurements (Figure \ref{fig:corr-log}). For scales larger than 10 $h^{-1}$ Mpc, this difference is within the 3-$\sigma$ errors. The disagreement for $s<1$ $h^{-1}$ Mpc is larger than 20\%. However, this can be due to the uncertainties introduced by the fiber collisions at those scales and effects of the resolution of the simulation. Therefore, we will focus our attention at scales larger than 5 $h^{-1}$ Mpc where the impact of fiber collision is smaller.

In order to study the clustering in different redshift bins using the HAM implemented in this work, we divide the full range into three bins. We select approximately the same number of galaxies in each redshift bin in order to have similar statistics in all of them. We perform an abundance matching (different scatter values that vary from 0.05 to 0.5) for each range to fit the monopole. Figure \ref{fig:quad-reds} shows the monopole and quadrupole for the three different redshift bins. The discrepancy in the quadrupole can be due to one or more of the approach used in this work. Possible causes of this discrepancy are enumerated below.

\begin{enumerate}
\item \citet{guo2015} find similar discrepancies in the quadrupole in configuration space for scales $>5$ $h^{-1}$ Mpc. They argue that the underestimation of the quadrupole on large scales is possible due to the correlated neighbouring bins in the covariance matrix. They obtain a reasonable $\chi^2$, even with this feature of the predicted quadrupole.

\item \citet{montero2014} show that the intermediate redshift bin ($0.51<z<0.57$) is the most complete region in the CMASS sample. The standard HAM can reproduce monopole and quadrupole for this redshift bin (see Figure \ref{fig:quad-reds}), but cannot reproduce the quadrupole for the other two bins. The CMASS DR12 sample has small variations in the monopole. However, quadrupole changes and it becomes similar for the two redshift ranges where the incompleteness of the sample is larger (Figure \ref{fig:quad-cmass}).
\begin{figure}
  \includegraphics[width=84mm]{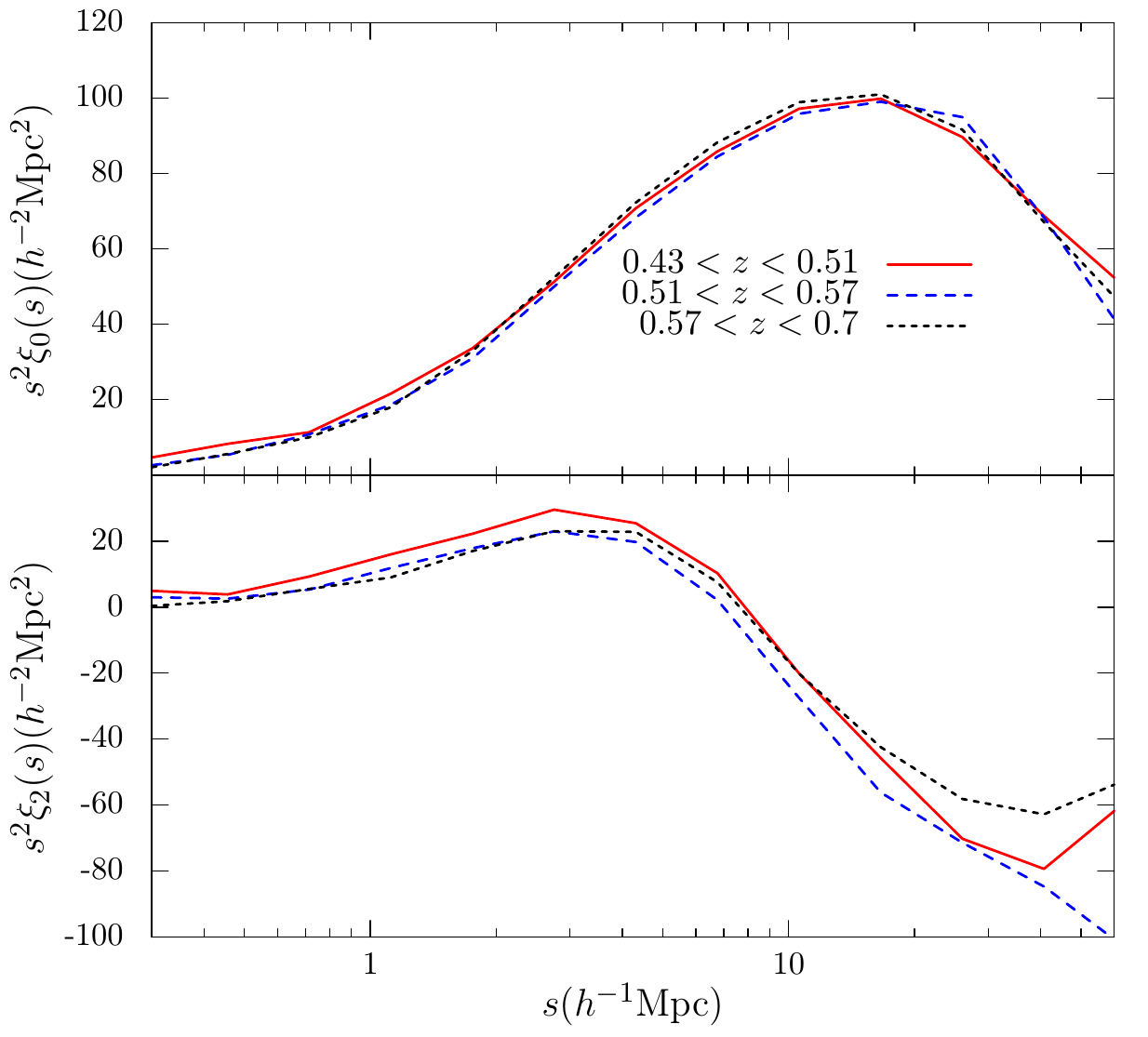}
  \caption{Correlation function for the CMASS sample in three redshift bins. $Top$ $Panel$: Monopole with small variations in time. $Bottom$ $Panel$: The quadrupole for the selected ranges. In contrast with the monopole, the quadrupole shows larger variations for  the different redshifts.}
  \label{fig:quad-cmass}
\end{figure}

\item The values of scatter used to fit the monopole of the correlation function in the different redshift bins vary in a wide range. This can be due to the evolution of the number density in the CMASS sample and some approximations used in this work. \citet{leauthaud2015} show a non-negligible evolution of the SMF at low redshift compared with the complete redshift range (0.43-0.7). Our approximation of non-evolving SMF could overestimate the incompleteness in the low redshift range (Figure \ref{fig:incomp}, left panel), then the necessary scatter to reproduce the observed correlation function will be smaller. We also assume a constant mean scatter, but indeed scatter depends on the stellar mass, it increases with the mass of the galaxies \citep{trujillo2011,reddick2013}. This dependency can explain why the scatter needed to reproduce the clustering of the low redshift range is smaller that the one used in the middle redshift. At low redshift the number density is equal to $3.466\times10^{-4}$ $h^3$ Mpc$^{-3}$, which is smaller than $3.942\times10^{-4}$ $h^3$ Mpc$^{-3}$ for the middle redshift. If both sample were complete, we will expect a larger scatter in the first range. However, due to the large incompleteness in the high mass end at low redshift, the mean mass of this sample is $1.86\times 10^{11}$ $M_\odot$ compared to $2.04\times 10^{11}$ $M_\odot$ for the second redshift range. For this reason, the scatter needed to reproduce the clustering is smaller in the low redshift range. In the high redshift bin, we can only see very massive galaxies (see Figure \ref{fig:incomp}, right panel) compared to the whole population of galaxies in the CMASS sample. This range is complete in the high mass end, and compared to the other two redshift ranges, has a number density very small ($1.534\times10^{-4}$ $h^3$ Mpc$^{-3}$), which implies larger mean mass ($2.63\times 10^{11}$ $M_\odot$) and scatter than for the other samples.

\item We have added a simple model for the stellar mass incompleteness in the CMASS sample. However, there can be other effects of the incompleteness in the target selection that cannot be modelled in this simple way. Although the selection is performed to select LRG, an incomplete blue cloud is in the sample and its fraction compared to the red sequence evolves with redshift  \citep[e.g.,][]{guo2013,montero2014}. Those two populations can live in different kinds of halos, and therefore they should be described by different scatter values. The errors introduced by this effect can increase with redshift, because the fraction of blue galaxies increases as well. As opposed to the low redshift bin, the high redshift bin is complete in the high-mass end (Figure \ref{fig:incomp}, $z=0.65$), but the fraction of blue galaxies is larger than the middle bin, which can affect the prediction of the quadrupole. The presence of a small fraction of the so-called ``green valley'' can also introduce small errors in our modelling.

\item The number density in the high redshift bin ($0.57<z<0.70$) is very small compared to the middle redshift range. In this region, the fraction of small galaxies decreases and the impact of the most massive objects in the clustering becomes stronger. \citet{guo2015b} and \citet{saito2015} need modification of the HAM model when colour cuts are applied. In addition, \citet{guo2015} show the necessity to introduce a velocity bias in the HOD to reproduce the most massive galaxies. If the standard HAM does not describe the clustering of the most massive galaxies, HAM mocks, which model samples as the CMASS in the redshift range $0.57<z<0.70$, will not reproduce accurately the clustering of the observed data.

\item In addition, recent papers reports results for Luminous Red Galaxies samples where the number of significant mis-central galaxies in halos is larger than expected \citep[e.g.,][]{hoshino2015} or the presence of off-centering for central galaxies \citep[e.g.,][]{hikage2013}. The implementation of these results in the construction of mocks reproducing LRG samples could also modify the quadrupole.
\end{enumerate}

\section{SUMMARY}
\label{sec:summary}

We investigated the galaxy clustering of the BOSS CMASS DR12 sample using light-cones constructed from the \textsc{BigMDPL} simulation. We perform a HAM to populate the dark matter halos with galaxies using the Portsmouth DR12 stellar mass catalogue. In addition, the stellar mass distribution is modelled to take into account the incompleteness in stellar mass of the CMASS sample. Our study included such features as the survey geometry, veto masks and fiber collision. The combination of HAM and the  \textsc{BigMDPlanck} simulation provides results in a good agreement with the observed data. Our results show that the HAM is a method extremely useful in the study of the relation between dark matter halos and galaxies, and can be very helpful in the production of mock catalogues \citep[][companion paper]{kitaura2015}. 

Our main results can be summarised as follows.
\begin{enumerate}
\item We model the observed monopole in configuration space using HAM. Assuming a complete sample, the scatter parameter is very large compared to previous studies. The modelling of stellar mass incompleteness significantly decreases the value of scatter to $\sigma_{\textsc{ham}}(V_{peak}|M_*)=0.31$. Our model reproduces the observed monopole for nearly every scale.
\item The prediction of the quadrupole in configuration space appears to be in disagreement with the observed data. We present possible explanations of this disagreement. In future works, we will concentrate in reduce the possible systematics, in order to understand better the limits of our model.
\item We compute the projected correlation function and the Three-point correlation function, finding a good agreement between the model and the observed data within 1-$\sigma$ errors for most of the scales. For scales $\sim0$ and $\sim\pi$, the differences are of the order of 2-$\sigma$ errors, which can be related to the same factors of the disagreement in the quadrupole. The monopole in $k$-space of the \textsc{BigMD-BOSS} light-cone is in a remarkable agreement with the measurement from the CMASS sample corrected by fiber collisions ($\sim$10\% of difference at $k=0.9$). The same agreement is found when we use nearest neighbour weights, which shows that the assignment of fiber collision in the light-cone can reproduce the observed data.
\item We compare our prediction of the stellar to halo mass relation with lensing measurements. The results are in a good agreement with the observed data. Our assumption of a constant scatter is reflected in the differences with observations. Lensing measurements suggest the need to include the stellar mass dependency in the scatter of the HAM.
\end{enumerate}

The \textsc{BigMD-BOSS} light-cone is publicly available. It can be found in the SDSS SkyServer\footnote{http://http://skyserver.sdss.org/dr12/en/home.aspx}. The current version includes angular coordinates (\textsc{ra, dec}), redshift in real space and redshift space, peculiar velocity in the line of sight, $M_{200}$, $V_{peak}$ and $M_*$. Properties of galaxies such as effective radius ($R_{eff}$), velocity dispersion ($\sigma_v$) and mass to light ratio ($M/L$) will be included in future updates.

\section*{Acknowledgements}

SRT is grateful for support from the Campus de Excelencia Internacional UAM/CSIC. SRT thanks Fernando Campos del Pozo for useful discussions and help while developing the \textsc{sugar} code.

The \textsc{BigMultiDark} simulations have been performed on the SuperMUC supercomputer at the Leibniz-Rechenzentrum (LRZ) in Munich, using the computing resources awarded to the PRACE project number 2012060963. The authors want to thank V. Springel for providing us with the optimised version of GADGET-2. 

SRT, CC, FP, AK, FSK, GF, and SG acknowledge support from the Spanish MICINNs Consolider-Ingenio 2010 Programme under grant MultiDark CSD2009-00064, MINECO Centro de Excelencia Severo Ochoa Programme under grant SEV-2012-0249, and grant AYA2014-60641-C2-1-P.  GY acknowledges support from MINECO (Spain) under research grants AYA2012-31101 and FPA2012-34694 and Consolider Ingenio SyeC CSD2007-0050. FP wish to thank the Lawrence Berkeley National Laboratory for the hospitality during the development of this work. FP acknowledges the Spanish MEC ``Salvador de Madariaga'' program, Ref. PRX14/00444.

CH also want to thank the Instituto de F\'isica Te\'orica UAM/CSIC for the hospitality during his summer visit, where part of this work was completed. GF acknowledges financial support from the Ministerio de Educaci\'on y Ciencia of the Spanish Government through FPI grant AYA2010-2131-C02-01. FSK acknowledges the support of the Karl-Schwarzschild Program from the Leibniz Society.

Funding for SDSS-III has been provided by the Alfred P. Sloan Foundation, the Participating Institutions, the National Science Foundation, and the U.S. Department of Energy Office of Science. The SDSS-III web site is http://www.sdss3.org/.

SDSS-III is managed by the Astrophysical Research Consortium for the Participating Institutions of the SDSS-III Collaboration including the University of Arizona, the Brazilian Participation Group, Brookhaven National Laboratory, Carnegie Mellon University, University of Florida, the French Participation Group, the German Participation Group, Harvard University, the Instituto de Astrof\'isica de Canarias, the Michigan State/Notre Dame/JINA Participation Group, Johns Hopkins University, Lawrence Berkeley National Laboratory, Max Planck Institute for Astrophysics, Max Planck Institute for Extraterrestrial Physics, New Mexico State University, New York University, Ohio State University, Pennsylvania State University, University of Portsmouth, Princeton University, the Spanish Participation Group, University of Tokyo, University of Utah, Vanderbilt University, University of Virginia, University of Washington, and Yale University

%%%%%%%%%%%%%%%%%%%%%%%%%%%%%%%%%%%%%%%%%%%%%%%%%%



\bibliographystyle{mnras}
\bibliography{references} % if your bibtex file is called example.bib


% Don't change these lines
\bsp	% typesetting comment
\label{lastpage}
\end{document}